\documentclass[10pt,twocolumn,showpacs,floatfix,sort]{revtex4-1}

\usepackage{textcomp}
\usepackage{epsfig}
\usepackage{dcolumn}
\usepackage{bm}
\usepackage[utf8]{inputenc}
\usepackage{amsmath}
\usepackage{amsfonts}
\usepackage{hyperref}
\newcommand{\field}[1]{\mathbb{#1}}

\usepackage{url}                         % For breaking URLs easily trough lines

\begin{document}

\title{Optimal reconstruction of dynamical systems: A noise amplification approach}

\author{L.\ C.\ Uzal}
\email{uzal@cifasis-conicet.gov.ar}
\author{G.\ L.\ Grinblat}
\email{grinblat@cifasis-conicet.gov.ar}
\author{P.\ F.\ Verdes}
\email{verdes@cifasis-conicet.gov.ar}
\affiliation{CIFASIS - French Argentine International Center for Information and Systems Sciences \\
UPCAM (France) / UNR - CONICET (Argentina) \\
Blvd.\ 27 de Febrero 210 Bis, S2000EZP Rosario, Argentina}

\begin{abstract}
In this work we propose an objective function to guide the search for a state space reconstruction of a dynamical system from a time series of measurements.  This statistics can be evaluated on any reconstructed attractor, thereby allowing a direct comparison among different approaches: (uniform or non-uniform) delay vectors, PCA, Legendre coordinates, etc.  It can also be used to select the most appropriate parameters of a reconstruction strategy.  In the case of delay coordinates this translates into finding the optimal delay time and embedding dimension from the absolute minimum of the advocated cost function.  Its definition is based on theoretical arguments on noise amplification, the complexity of the reconstructed attractor and a direct measure of local stretch which constitutes a novel irrelevance measure.  The proposed method is demonstrated on synthetic and experimental time series.
\end{abstract}


\maketitle

\section{Introduction}

\subsection{Background}

Nonlinear time series analysis has been actively developed during the last decades following a fundamental theorem by Takens \cite{Takens:1981}.
He proved that it is possible to reconstruct the attractor of a dynamical system using only a single time sequence of scalar measurements from the system under study.
The vectors accomplishing this reconstruction have the form $\bar{x}(t) = (x(t), x(t-\tau),..., x(t-(m-1)\tau))$, where $x(t)$ are our observations on the system, the integer number $m$ is known as the embedding dimension, and $\tau$ is the time difference between consecutive components, also called time lag or delay time, which is some multiple of the sampling time.
Takens assumed an infinite sequence of noise-free measurements and proved the existence of a diffeomorphism between the original and reconstructed attractors for almost any choice of positive delay times $\tau$ and a sufficiently big dimension $m$.  This approach is known as
{\it delay coordinate embedding} and constitutes the first step of almost all nonlinear time series analysis methods such as the determination of attractor dimensions, Lyapunov exponents and entropy \cite{Kantz:1997}.
Takens' ideas were later revisited by Sauer {\it et al}.\ \cite{Sauer:1991} who proved that for compact attractors the embedding dimension $m$ must be larger than twice the box-counting dimension of the original attractor to ensure a one-to-one reconstruction. In practice, however, the box counting dimension of the original attractor is unknown and as a consequence the minimal embedding dimension must be derived from the data.

The reconstruction problem starts with the measurement of appropriate physical quantities in order to ensure further system state reconstruction. This is a very important instance in the reconstruction process which affects the rest of the process.  The problem of selection of an optimal measurement function allowing an information flow from the unobserved variables to the observed variable was first discussed in \cite{Casdagli:1991} and later studied in \cite{Letellier:1998,Smirnov:2002,Letellier:2005}.  This is a very interesting subject but out of the scope of this paper.

The selection of the delay time $\tau$, although irrelevant in Takens' formal derivation, becomes important for experimental time series due to their finite length and measurement precision.
More precisely, the quality of the reconstruction and the extraction of diffeomorphic invariants thereof are affected by the time window size $t_w=(m-1)\tau$ determined by the choice of $\tau$ and $m$ as argued in the literature by many authors \cite{Casdagli:1991, Gibson:1992, Albano:1991, Martinerie:1992, Potapov:1997, Kember:1993, Kugiumtzis:1996, Kim:1998, Kim:1999, Small:2004}.
If a small (large) value of $t_w$ is chosen a phenomenon known as redundance (irrelevance) deteriorates the reconstruction quality \cite{Casdagli:1991, Gibson:1992}.
We therefore see that, for delay coordinate reconstructions, two parameters (say $\tau$ and $m$ or $t_w$ and $m$) need to be chosen according to some optimality criteria.  In general, it can be expected that the optimal reconstruction parameter values will be determined simultaneously.

Many authors have proposed methods to select an optimal delay time by minimizing redundance between components \cite{Albano:1991, Fraser:1986, Liebert:1989, Fraser:1989a, Buzug:1992a, Kember:1993, Rosenstein:1994, Aguirre:1995, Kim:1998, Kim:1999}.
However, these methods require additional argumentations on how to avoid irrelevance.
In particular, Fraser and Swinney proposed, in their early work \cite{Fraser:1986}, to use the first minimum of the Mutual Information (MI) between delayed components, and their method has now become the reference standard.
By using the first instead of the absolute minimum they attempted to bias the selection towards small delays, because large ones could imply irrelevance between components.
Such criterion has potentially three drawbacks:
i) The first minimum could be a noise-induced fluctuation instead of a true local minimum \cite{Martinerie:1992}.
ii) There is in general no evidence that, in addition to minimizing redundance, the ad-hoc choice of the {\it first} relative minimum of MI should as well minimize irrelevance, or at least keep it low.
iii) Most chaotic flows have a characteristic oscillating period which will modulate the MI profile generating a structure of maxima and minima.
More precisely, the first maximum of MI will occur at this characteristic period, which will therefore act as an upper bound for the first minimum even if the irrelevance time of the time series is larger than this period.
These three arguments are valid for all methods in \cite{Fraser:1986,Liebert:1989, Fraser:1989a, Buzug:1992a, Kember:1993, Rosenstein:1994, Aguirre:1995, Kim:1998, Kim:1999}, where a redundance measure is proposed and a first minimum (or maximum) must be determined.

\begin{figure}
\centering
\includegraphics[width=0.48\textwidth]{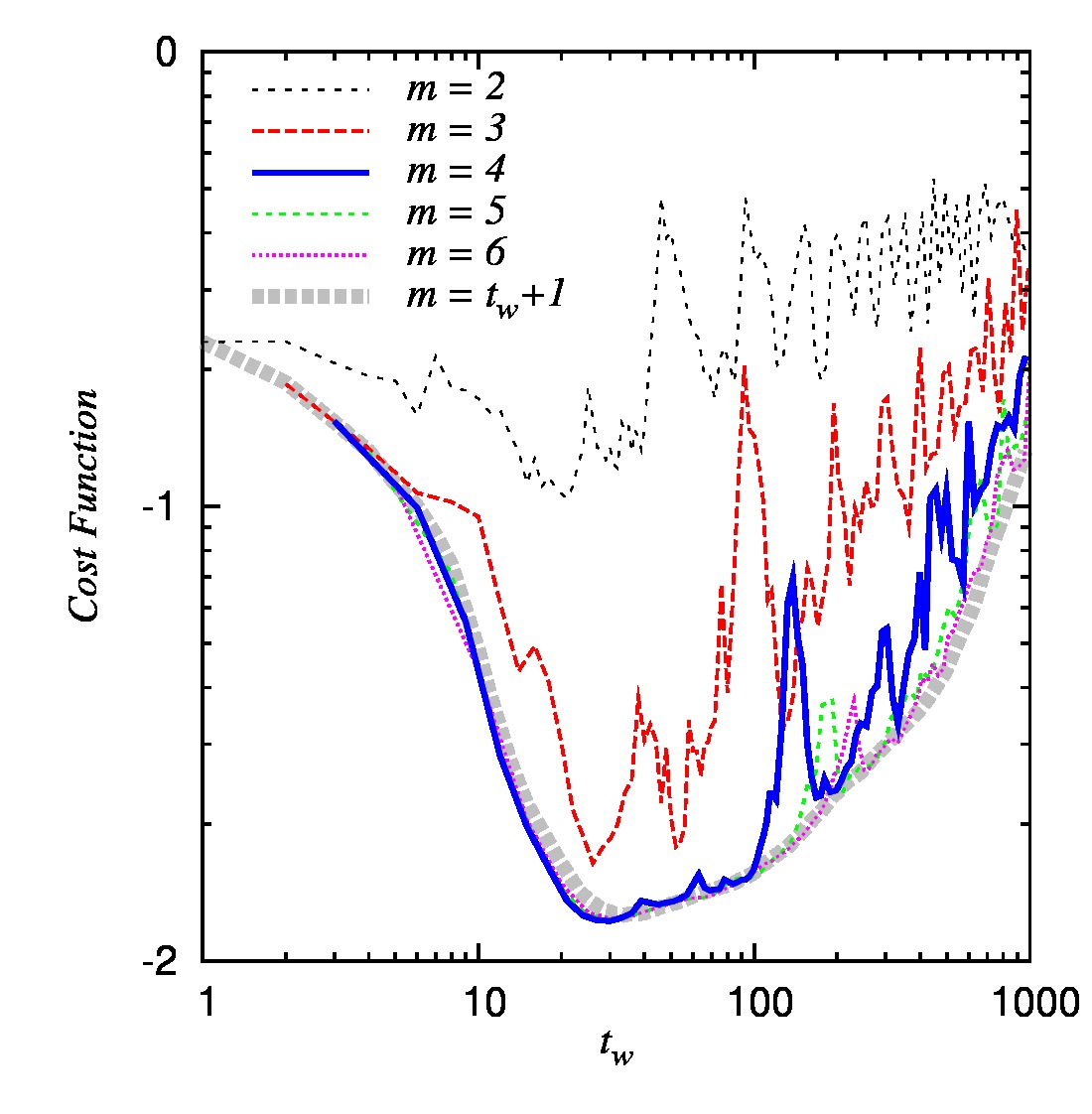}
\caption{Profiles of the proposed cost function as a function of the dimension $m$ and time window $t_w$ for the Mackey-Glass time series. The thick dashed line corresponds to the case of using all delay coordinates within the time window $t_w$ (i.e. using $\tau=1$ and $m=t_w+1$, with $\tau$ and $t_w$ expressed in units of sampling time). The cost function has an absolute minimum at $m=4$, $t_w=30$.}
\label{fig:example_mcgl}
\end{figure}

On the other hand, methodologies based on dynamical arguments can be found in the literature on attractor reconstruction \cite{Buzug:1992a, Liebert:1991, Gao:1993, Kennel:2002} which can simultaneously determine both the delay time and the embedding dimension.
These methodologies are based on tracking the dynamical evolution of nearest neighbors in the reconstructed space and measuring their divergence, and are intimately related to the concept of False Nearest Neighbors (FNN) introduced by Kennel {\it et al}.\ \cite{Kennel:1992}.
In this manuscript we will refer to them as {\it dynamical methods}.
The idea of seeking false neighbors to determine the embedding dimension was considered also in \cite{Cenys:1988,Aleksic:1991,Cao:1997}.

Recent years have seen an increasing interest on the development of reconstruction techniques for prediction purposes \cite{Pi:1994, Judd:1998, Tsui:2002, Small:2004, Hirata:2006, Simon:2007, Manabe:2007, Tikka:2008, Ragulskis:2009, Holstein:2009, Lukoseviciute:2010}.
These recent studies have focused on producing {\it non-uniform delay coordinate} vectors whereby each component is associated to a different delay time.
The problem of non-uniform delay reconstruction has also been addressed by Pecora {\it et al.} \cite{Pecora:2007} and Garcia {\it et al.} \cite{Garcia:2005a,Garcia:2005b}.
Both methods require the selection of the first minimum (or maximum) of their proposed measures and are only applicable to (non-uniform) delay coordinate reconstructions.
Holstein and Kantz \cite{Holstein:2009} proposed a generalized embedding approach for the case of time series modeling in a Markovian sense.

The quality of a reconstruction was quantified by Casdagli {\it et al}.\ \cite{Casdagli:1991} in terms of its (observational) noise amplification effect when one wishes to estimate the state of the system. They defined the {\it noise amplification} $\sigma$ to locally measure this effect, a statistics which can potentially be computed from the observed time series.
In principle, $\sigma$ allows for an absolute comparison among reconstructions, which in turn enables a selection of optimal reconstruction parameters.
An obstacle, however, is given by the hypothesis of a full knowledge on the true generating dynamics.
In \cite{Casdagli:1991} the authors suggest to estimate $\sigma$ via a data-driven approximation of the dynamical evolution law.

In this work we build on the idea of noise amplification of Casdagli {\it et al.}\ to define a new cost function which is fully computable from the time series and its reconstruction.
Embedding parameters are then determined from the absolute minimum of the proposed objective function, which can be calculated for any kind of reconstruction.

\subsection{Overview of this work}

In this paper we propose a criterion to select an optimal state-space reconstruction of the dynamics of a physical system from a time series of measurements
performed on the system of interest.
The presented methodology is based on the minimization of a cost function $L$ which is readily computable from the available observational data.
Different reconstructions, whether multivariate or time-delayed univariate, uniform (equally spaced) or not, can be directly compared through $L$, and thereby the suitability of different reconstruction settings can be assessed.

For example, Fig.\ \ref{fig:example_mcgl} illustrates how optimal parameter values for the dimension $m$ and time window $t_w$ can be simultaneously determined by a global optimization of the proposed cost function for the Mackey-Glass time series \cite{Mackey:1997}.
An important advantage of the proposed approach is given by its automatic and objective character, in contrast to, for example, the subjective or practitioner-dependent choice on the location of the first local minimum of Mutual Information (MI), or the value of a threshold characterizing a negligible fraction of false nearest neighbors (FNN).

\begin{figure*}
\centering
\includegraphics[width=\textwidth]{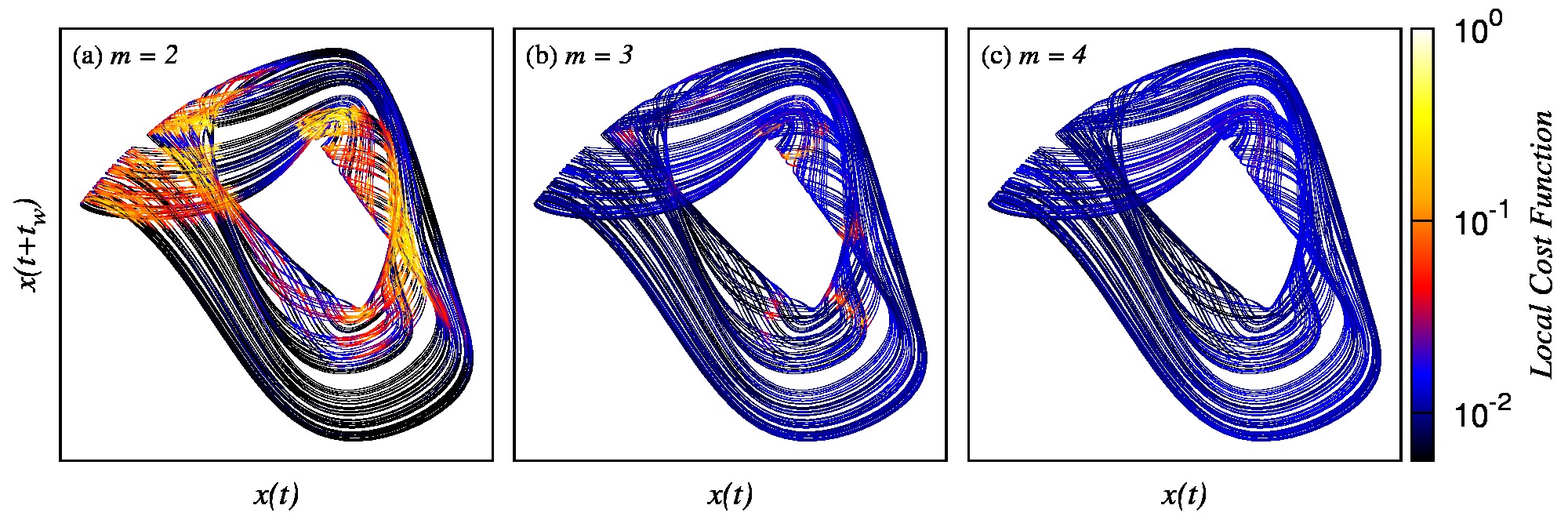}
\caption{(Color online). Color-coded local cost function in two-dimensional projections of $m$-dimensional delay coordinate reconstructions of the Mackey-Glass attractor. The time window $t_w=30$ is the same for all reconstructions but the dimension increases from left to right. (a) $m=2$. (b) $m=3$. (c) $m=4$.}
\label{fig:color}
\end{figure*}

The proposed cost function $L$ is a local property of the reconstruction, which we then average over the attractor. In Fig.\ \ref{fig:color} we illustrate the local behavior of $L$ by plotting a two-dimensional projection of the Mackey-Glass attractor \cite{Mackey:1997} as we reconstruct it in spaces of increasing dimension ---from $m=2$ (panel (a)) to $m=4$ (panel (c)). As we can see in panel (a), $L$ correctly senses regions of the attractor not yet unfolded for $m=2$, i.e.\ regions where orbits with different dynamical evolutions overlap. These regions progressively vanish in higher-dimensional reconstructions, as reflected by lower values of $L$ in panels (b) and (c).
Notice that the local cost function serves as a complementary tool that enables an additional intuition for the problem: the global cost function difference between the $m=3$ and the optimal reconstruction ($m=4$, Fig. \ref{fig:example_mcgl}) gains significance in the light of Fig.\ \ref{fig:color}(b) which shows localized regions of the attractor with large values of the local cost function.

\begin{figure}
\centering
\includegraphics[width=0.48\textwidth]{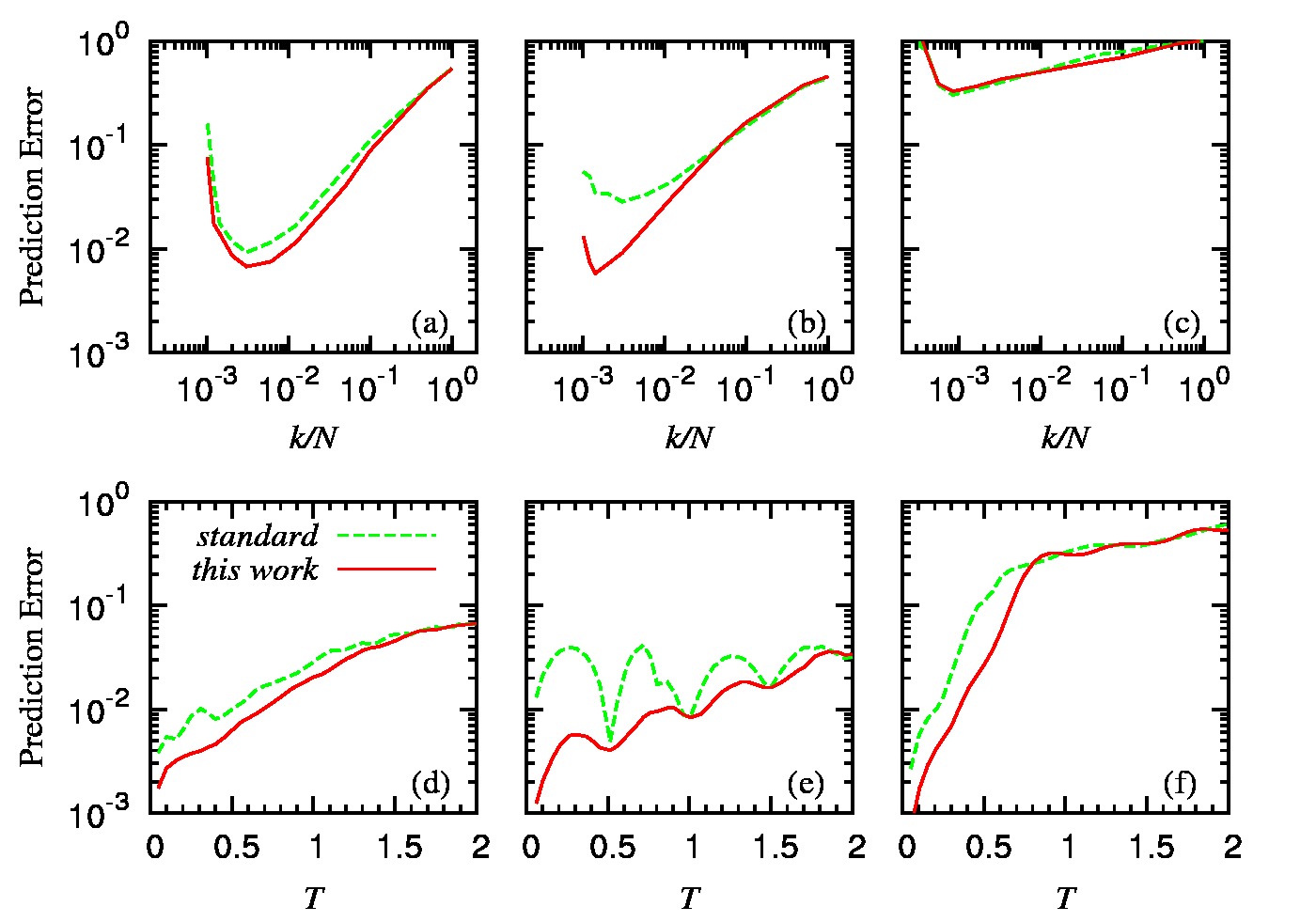}
\caption{(Color online). Comparison of prediction errors (normalized root mean squared error) of local linear models for reconstructions obtained following the standard approach (MI + FNN, dashed line) versus the proposed methodology (full line) for Mackey-Glass (panels {\it a} and {\it d}), R\"ossler ({\it b}, {\it e}) and Lorenz ({\it c}, {\it f}).  Top panels: forecasting error at a fixed horizon as a function of the database fraction ($k/N$) employed for model building.  Bottom panels: prediction error at a range of horizons $T$ from short to long for a fixed number of neighbors. The horizon $T$ is measured in units of the first recurrence time.}
\label{fig:validation}
\end{figure}

As an independent validation of the proposed reconstruction methodology we here consider forecasting performance for a range of horizons from short to long term.  More precisely, we use the proposed approach to determine a reconstruction space and compare the prediction accuracy of local linear models against the reconstruction arrived at following the standard approach (MI + FNN).  The top panels of Fig.\ \ref{fig:validation} contrast prediction errors at a fixed horizon as a function of the number of neighbors $k$ involved in the prediction for Mackey-Glass, R\"ossler and Lorenz datasets (for a more detailed account of the datasets and simulation settings employed please refer to Appendix \ref{ap:prediction}).  The lower panels, instead, illustrate the case of a fixed number of neighbors and a variable prediction horizon $T$ (for ease of interpretation, measured in units of the characteristic ---first recurrence--- time of each system).  Figure \ref{fig:validation} demonstrates that more accurate forecasting can be achieved at a range of horizons with the proposed approach.  This follows from the fact that our methodology produces smooth embeddings for which the complexity of the prediction law is minimized and the dynamics can be more efficiently approximated, as discussed in the following sections.

As we detail below, $L$ is built upon the concept of noise amplification introduced by Casdagli {\it et al}.\ \cite{Casdagli:1991}.
However, in this manuscript we present a new interpretation of $\sigma$ in terms of the complexity of the prediction law defined on the reconstructed attractor, as advanced in the previous paragraph.
We also show that the practical implementation of the proposed method is strongly related to: a) the FNN (False Nearest Neighbors) method proposed by Kennel {\it et al}.\ \cite{Kennel:1992}, and b) dynamical methods \cite{Buzug:1992a, Liebert:1991, Gao:1993, Kennel:2002}.
The cost function also incorporates a novel irrelevance measure based on a direct computation of the reconstructed attractor local stretch.

In Section \ref{sec:recprob} we introduce the notation we shall use throughout this work, classify the most common reconstruction strategies and discuss under which conditions a reconstruction is optimal.
We review the definition of noise amplification in Sections \ref{sec:noiseamp}.
We then show the equivalence between this definition and a measure of complexity of the resulting prediction law,
and discuss the consequences of this reinterpretation in Section \ref{sec:complexity}.
Then, in Section \ref{sec:sigma_k} we propose an estimation method based on a first neighbors approximation which enables the computation of noise amplification from time series.
In Section \ref{sec:norm} we propose a formulation of the cost function which naturally incorporates a penalization term to account for irrelevant components.
In Section \ref{sec:related} we discuss how our approach is related to other methods in the literature.
In Section \ref{sec:application} we present field data application examples.
Finally, in Section \ref{sec:conclusions} we draw our conclusions.

\section{The reconstruction problem}\label{sec:recprob}

\subsection{Introduction and notation}\label{sec:introduction}

In this work we will follow the notation employed in \cite{Casdagli:1991}.
The time series $x(t)$ is the sequence of measurements performed on the system under study at regular intervals $\delta t$.
We indicate with $s(t)$ the state of system at time $t$, which evolves according to a deterministic dynamics on a $d$-dimensional manifold $M$:
\begin{equation}
s(t)=f^t(s(0)),
\end{equation}
where $f^t$ is the evolution law for a time step $t$.
The time series is then the sequence $x(t)=h(s(t))$, where $h:M\rightarrow \field{R}$ is a smooth measurement function defined on the original state space.
The time series $x(t)$ is the only observable, being $s(t)$, $f^t$ and $h$ unobservables locked in a black box \cite{Casdagli:1991}.

\begin{figure}
\centering
\includegraphics[width=0.48\textwidth]{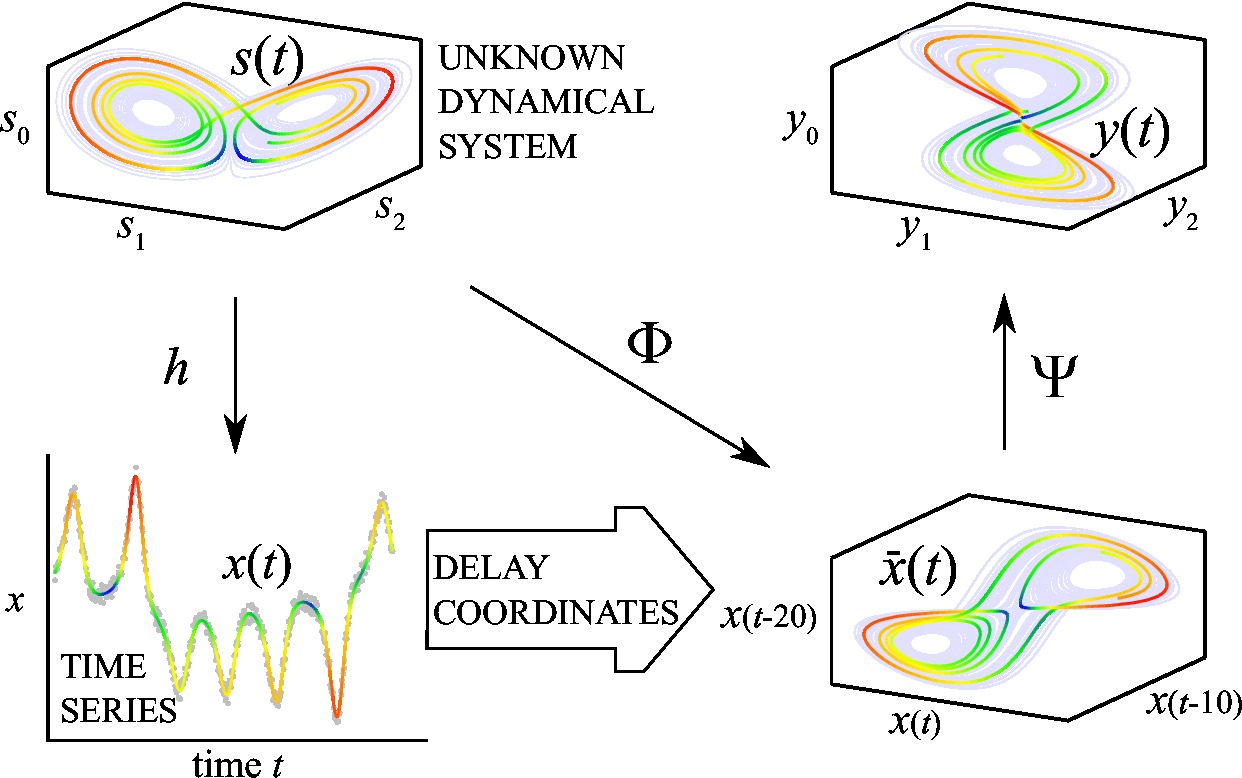}
\caption{(Color online). Schematic representation of the measurement process and reconstruction of the attractor of a dynamical system. This figure is inspired in Casdagli {\it et al.}\ \cite{Casdagli:1991}. See the main text for details.}
\label{fig:rec}
\end{figure}

Figure \ref{fig:rec} gives a schematic representation of the state space reconstruction process, starting from the measurement process.
In its most general form, a reconstruction of the state space vector $s(t)$ can be described with an $n$-dimensional vector $y(t)=\varPsi(\bar{x}(t))$, where $\bar{x}(t) = (x(t), x(t-\tau),..., x(t-(m-1)\tau))$ is the delay coordinate (DC) vector at time $t$, and $\varPsi:\field{R}^m\rightarrow \field{R}^n$ is a further transformation that accounts for the possibility of considering a more general transformation (e.g. non-uniform delay coordinates, global or local SVD, or a noise reduction algorithm).
Here we will first focus on DC reconstructions $\bar{x}(t)$, but the final methodology will be generally valid for arbitrary reconstructions of the form $y(t)$ described above.

The DC reconstruction defines a map $\varPhi:M\rightarrow \field{R}^m$ such that $\bar{x}=\varPhi(s)$. Under the hypotheses considered by Takens, $\varPhi$ will yield a diffeomorphism, and in such a case we will obtain an embedding of the attractor in the reconstructed space.
The dynamics is given by a function $F^t$ fully determined by the original dynamical law $f^t$ and $\varPhi$:
\begin{equation}\label{eq:evol_F}
\bar{x}(t)= F^t(\bar{x}(0))=\varPhi \circ f^t \circ\varPhi^{-1}(\bar{x}(0)).
\end{equation}

The value of $x(t+T)$ will be given by the reconstructed vector at time $t$, and can be obtained by applying the evolution operator $F^T$ to $\bar{x}(t)$ and retaining its first component.
This operation can be captured by the function
\begin{equation}
g^T=\pi\cdot F^T
\end{equation}
where $\pi$ is the column vector $(1, 0,\ldots, 0)$, in such a way that $x(t+T)=g^T(\bar{x}(t))$.
The same as $F^T$, the function $g^T$ is completely determined by $\varPhi$, the original evolution law $f^T$, and the measurement function $h$, being
\begin{equation}\label{eq:gT}
g^T=h\circ f^T \circ\varPhi^{-1}.
\end{equation}

In the following we will use the simplified notation $x(T)$ instead of $x(t+T)$ to denote the value of the time series $T$ time steps after $t$, which will be implicit in the notation.  Accordingly, $\bar{x}$ and $s$ will refer to $\bar{x}(t)$ and $s(t)$ respectively.

\subsection{Reconstruction strategies}

Delay coordinate (DC) reconstruction is the most common strategy for attractor reconstruction.  However, several alternatives exist ranging from derivative coordinates \cite{Packard:1980} or global principal value decomposition \cite{Broomhead:1986} to {\it non-uniform} DC vectors. The latter possibility has been extensively explored in the literature in recent years \cite{Pi:1994, Judd:1998, Tsui:2002, Small:2004, Garcia:2005a, Garcia:2005b, Hirata:2006, Pecora:2007, Huke:2007, Simon:2007, Manabe:2007, Tikka:2008, Ragulskis:2009, Holstein:2009, Lukoseviciute:2010}.
All of these approaches can be described in terms of Fig.\ \ref{fig:rec}, i.e. by means of a transformation $\varPsi$ applied to the DC vector.
Indeed, any coordinates defining a reconstruction $y(t)$ from samples in the interval $[t-t_w,t]$ can be thought of as the result of applying a transformation $\varPsi$ to the {\it full delay coordinate vector} (fDC) which contains all available data from $[t-t_w,t]$ (i.e.\ $\tau=1$ and $m=t_w+1$, with $\tau$ and $t_w$ expressed in units of sampling time).

Back on the domain of linear transformations, Gibson {\it et al}.\ \cite{Gibson:1992} found an analytical solution for PCA in the limit of a small time window width.
The `small window regime' refers to widths smaller than
\begin{equation}\label{eq:tauCasdagli}
\tau^*_w = 2 \sqrt{3 \langle x^2 \rangle / \langle (dx/dt)^2 \rangle}
\end{equation}
where the time series values $x$ have been normalized to zero mean.
Their analytical solution has principal directions given by discrete Legendre polynomials.
For this case, $\varPsi$ is a linear transformation which projects the fDC vector from the time window $t_w$ onto $n$ directions given by the first $n$ discrete Legendre polynomials.
Therefore, the free parameters to build the Legendre coordinates are the time window width $t_w$ and the dimension $n$.
Gibson {\it et al.}\ gave a guidance for choosing $t_w$, but they argued that there is no simple rule to select $n$.
In order to avoid redundance and irrelevance and achieve a good balance between signal-to-noise ratio and complexity, they proposed to use a window width {\it smaller than but close to $\tau^*_w$} \cite{Gibson:1992}.
However, there is no demonstrated relationship between $\tau^*_w$ and the characteristic irrelevance time, and it can not be discarded that time window widths larger than $\tau^*_w$ could provide useful information about the system state for its reconstruction.

Gibson {\it et al.}\ also showed that, in the limit of a small DC dimension $m$, applying the discrete Legendre polynomials transformation to the fDC vector is equivalent to a finite differencing filter which will recover derivative coordinates.  Therefore, derivative coordinates are also encompassed by the same linear transformation framework.

Finally, {\it non-uniform delay coordinate} (nuDC) reconstruction has been proposed as a generalization of standard DC reconstruction by allowing a different time lag for each component of the DC vector, namely
\begin{equation}\label{eq:nonuniform}
y(t) = (x(t), x(t-\tau_1),x(t-\tau_2),..., x(t-\tau_{(n-1)})),
\end{equation}
where the dimension $n$ and the $n-1$ delay times $\{\tau_1,\tau_2,...,\tau_{(n-1)}\}$ are the set of parameters to be determined.
Also in this case the reconstructed vector $y$ can be obtained from the fDC vector on $t_w=\max_i(\tau_i)$.
The dimension is reduced from $m=t_w+1$ to $n$ by retaining only the $n$ coordinates corresponding to the non-uniform delays.
It is important to remark that this procedure constitutes a linear projection $\varPsi$ onto a subspace and that the iDC reconstruction is not a new procedure outside the reconstruction scheme of Fig.\ \ref{fig:rec} proposed in \cite{Casdagli:1991}.

In summary, existing strategies in the literature consist of a linear projection of the fDC vector onto a subspace (in particular, this is also the case for the usual, uniform DC reconstruction).
It is unclear which strategy will be optimal in each case.

\subsection{When is the reconstruction optimal?}\label{sec:whenisoptimal}

Once the reconstruction problem has been defined, the question arises as how to choose the free parameters $t_w$ and $m$ (and, eventually, also the parameters associated to a further transformation $\varPsi$) in order to achieve the best possible reconstruction.
To answer this question one must define the criterion of optimality.
In other words: what would {\it best} mean for the purpose of the reconstruction?
Takens' theorem gives no guidance to define such criterion, being almost any $\tau$ and a big enough $m$ equally valid solutions for the noise free and infinite amount of data case analyzed by the theorem.
The presence of noise and the finite amount of data introduce limits to the quality of the reconstruction in the sense that any measure we would later estimate from it would require modeling the distribution of points, introducing noise amplification and estimation error \cite{Casdagli:1991}.
The aim of the optimal reconstruction will be to simultaneously minimize these two effects.
Observational noise amplification is reduced for well unfolded attractors in the reconstructed space because, roughly speaking, well separated states will be harder to mix by noise displacements.
On the other hand, excessive unfolding will at some point produce an overly complicated reconstruction that will later require more data points in order to be modeled.
The compromise between these two opposite scenarios will then depend on the noise level and the amount of available data.

Another idea of optimality is based on minimizing the distortion of the original attractor introduced when applying the reconstruction map $\varPhi$.
If we assume that the original attractor is known, then it is possible to define measures of distortion between the two attractors as in \cite{Fraser:1989b,Casdagli:1991,Potapov:1997}.
However, there is no reason to assume that the original attractor constitutes the best representation of itself.
For example, Pecora {\it et al.} suggest that a combination of original and delay coordinates of the Lorenz attractor can produce a better reconstruction than the original attractor itself \cite{Pecora:2007}.
Beyond these arguments, we will not use these distortion measures as our cost function because we will assume that the original attractor is unknown.
However, the point we would like to question here is whether these measures \cite{Fraser:1989b,Casdagli:1991,Potapov:1997} are absolute measures of the reconstruction quality.

\section{Construction of the cost function}\label{sec:defining}

\subsection{Noise amplification}\label{sec:noiseamp}

The concept of noise amplification as given by Casdagli {\it et al}. \cite{Casdagli:1991} aims at quantifying the effect that observational noise on $x$ has on our uncertainty about the system state $s$.
Given that the state of the system is unknown and we only have access to the observational time series $x=h(s)$, it is impossible to evaluate the quality of a reconstruction by comparing the reconstructed attractor with the original one.
However, the quality of a reconstruction can be assessed by considering the predictive power that it allows for.
In this context, the definition of noise amplification (see \cite{Casdagli:1991} for details) is given by:
\begin{equation}\label{eq:sigma}
\sigma(T,\bar{x})=\lim_{\epsilon\rightarrow 0}\sigma_\epsilon(T,\bar{x})
\end{equation}
where
\begin{equation}\label{eq:sigmaeps}
\sigma_\epsilon(T,\bar{x})=\frac{1}{\epsilon}\sqrt{Var(x(T)|B_\epsilon(\bar{x}))}
\end{equation}
being ${Var(x(T)|B_\epsilon(\bar{x}))}$ the conditional variance of $x(T)$ for $\bar{x}$ in a radius $\epsilon$ ball $B_\epsilon(\bar{x})$ defined by an observational noise level $\epsilon$.
According to the definition given in \cite{Casdagli:1991}, ${Var(x(T)|B_\epsilon(\bar{x}))}$ does not contain any contribution from modeling error. Instead, the exact form of $g^T$ is assumed to be known and used to compute the width of the image of $B_\epsilon(\bar{x})$ when passed through $g^T$.
In the limit $\epsilon\rightarrow 0$ the value of $\sigma(T,\bar{x})$ is independent from $\epsilon$ but only a function of the reconstruction as given by $\varPhi$.
Finally, in \cite{Casdagli:1991} the authors considered the mean square value of $\sigma(T,\bar{x})$ with respect to the measure of the attractor, $\langle\sigma^2(T)\rangle$, to globally characterize the reconstruction.

The forward time step $T$ on which ${Var(x(T)|B_\epsilon(\bar{x}))}$ is evaluated is a free parameter of $\sigma(T,\bar{x})$.
Notice, however, that the evaluation of $\sigma(T,\bar{x})$ on a single value of $T$ is insufficient to correctly characterize the
divergence of reconstructed orbits since the measurement function can collapse different states onto the same value.
A more robust measure of the divergence between neighboring orbits is obtained by considering the average of $\sigma^2(T,\bar{x})$ for $T$ on the interval $[0, T_M]$ for some upper value $T_M$.

Therefore, we redefine $\sigma_\epsilon(\bar{x})$ as
\begin{equation}\label{eq:E}
\sigma_\epsilon^2(\bar{x})=\frac{1}{T_M}\int_0^{T_M}{\sigma_\epsilon^2(T,\bar{x})dT},
\end{equation}
and consider the limit
\begin{equation}\label{eq:barsigma}
\sigma(\bar{x})=\lim_{\epsilon\rightarrow 0}\sigma_\epsilon(\bar{x})
\end{equation}
instead of the original definitions given in eqs.\ (\ref{eq:sigma}) and (\ref{eq:sigmaeps}) respectively.
We will keep the notation $\sigma_\epsilon(T,\bar{x})$ (with an explicit indication of parameter $T$) for the cases where the average over $T$ in $[0,T_M]$ is not performed and when we would like to evaluate this quantity \----and others derived from it\---- at a single instance in future, $T$.

The choice of parameter $T_M$ will determine the sensitivity of $\sigma(\bar{x})$ to the quality of the reconstruction.
However, there is no need to make an accurate selection of $T_M$ provided that it is large enough as to capture orbits divergence.
The purpose of setting an upper bound for $T_M$ is just to achieve a better sensitivity of $\sigma$ to the optimum reconstruction.
Our method based on the estimation and minimization of $\sigma_\epsilon(\bar{x})$ produces consistent results for a wide range of $T_M$ values, as we shall show in Section \ref{sec:related_dynamical}.

\subsection{Complexity of the prediction law}\label{sec:complexity}

Although the purpose of the definition of $\sigma(T)$ is to characterize the reconstruction determined by $\varPhi$, it depends exclusively on the function $g^T$.  Therefore, it should be possible to derive an expression of $\sigma(T)$ in terms of $g^T$ only.  This is the purpose of this section.

Let $B_\epsilon(\bar{x})$ be a Gaussian ball with standard deviation $\epsilon$, i.e.\ a multivariate normal distribution with a diagonal covariance matrix $\Sigma$ with all entries equal to $\epsilon^2$. In this context, ${Var(x(T)|B_\epsilon(\bar{x}))}$ is the variance of this Gaussian distribution mapped through $g^T$ to $\field{R}$. As we aim to take the limit $\epsilon\rightarrow 0$, we can consider $\epsilon$ small enough as to make a first order approximation around $\bar{x}$
\begin{equation}\label{eq:firstorderg}
g^T( \bar{x}+\xi)=g^T( \bar{x})+b^\dagger\xi + \mathcal{O}(\|\xi\|^2)
\end{equation}
where $b= \nabla g^T(\bar{x})$ and $\xi$ represents a displacement from $\bar x$ with a size $\|\xi\|$ bounded by $\epsilon$.
In this linear limit the resulting mapped distribution is also normal with a variance given by $b^\dagger \Sigma b=\epsilon^2\|\nabla g^T(\bar{x})\|^2$.
According to this result we find that
\begin{equation}\label{eq:grad}
\sigma^2(T,\bar{x})=\|\nabla g^T(\bar{x})\|^2.
\end{equation}
This relationship, not considered in \cite{Casdagli:1991}, allows for a new interpretation of the minimization of $\langle\sigma^2(T)\rangle$ in terms of the complexity of the resulting $g^T$.
A change in the reconstruction modifies the support where $g^T$ lives and therefore $g^T$ itself.
The smoothness of $g^T$ can be quantified by $\langle\|\nabla g^T(\bar{x})\|^2\rangle$, which can then be interpreted as a measure of the complexity of this function. Minimizing $\langle\sigma^2(T)\rangle$ is therefore equivalent to minimizing the complexity of $g^T$.

This new interpretation is per se sufficient to consider the minimization of $\langle\sigma^2(T)\rangle$ as an optimality criterion to guide the reconstruction process, independently of the original arguments on the effects of observational noise.
The law $g^T$ will be easier to model the lower the value of $\langle\sigma^2(T)\rangle$ is, and therefore fewer parameters will be needed to describe it.
Furthermore, smoothness of the dynamics is the first assumption when modeling a system with the purpose of forecasting ---e.g.\ with techniques such as artificial neural networks, radial basis functions, or interpolating splines. Building on this hypothesis, all of
these modeling approaches include a regularization parameter to avoid overfitting \cite{Hastie:2001}. More importantly,
even when forecasting is not the final purpose but estimating an invariant quantity such as the maximum Lyapunov exponent, these algorithms are also implicitly applied, being $k$-NN the most frequently used.
Therefore, it is crucial that the unknown function $g^T$ complies, as closely as possible, with the hypotheses made on it when it is modeled.

A further consequence of this reinterpretation is the possibility to generalize the definition of $\sigma$ to the more general reconstruction vector $y=\varPsi(\bar{x})$.
In \cite{Casdagli:1991} the $\epsilon$-ball $B_\epsilon(\bar{x})$ is originated in i.i.d.\ observational noise in each component of the DC vector $\bar{x}$ and can only be associated to it.
Any further transformation $\varPsi$ will distort this noise-induced $\epsilon$-ball.
For the new interpretation in terms of $\|\nabla g^T(\bar{x})\|^2$, the $\epsilon$-ball is a mathematical construction to capture the behavior of the neighborhood of $\bar{x}$, and as such is therefore also applicable to $y$.

\subsection{Estimation of noise amplification with $k$-NN}\label{sec:sigma_k}

In practical applications the law $g^T$ is inaccessible (being unknown both the dynamical law $f^t$ and the measurement function $h$), and the only available information is the time series itself.
In this context we propose to estimate $\sigma_\epsilon^2(\bar{x})$ by recursing to the nearest $k$ neighbors of $\bar{x}$ \cite{comment:Theiler}.
These $k$ neighbors and $\bar{x}$ define a set $\mathcal{U}_k(\bar{x})$ with $k+1$ elements which will act as a proxy for the ball $B_\epsilon(\bar{x})$ in the following definitions.

We approximate the conditional variance ${Var(x(T)|B_\epsilon(\bar{x}))}$ by
\begin{equation}\label{eq:Vark}
E_k^2(T,\bar{x})\equiv\frac{1}{k+1}\sum_{{\bar{x}}'\in\mathcal{U}_k(\bar{x})}[x'(T)-u_k(T,\bar{x})]^2
\end{equation}
where $x'(T)$ is the future value of $x$ corresponding to $\bar{x}'$, and
\begin{equation}
u_k(T,\bar{x})\equiv\frac{1}{k+1}\sum_{\bar{x}'\in\mathcal{U}_k(\bar{x})} x'(T).
\end{equation}

In order to capture the time average over $T$ in $ [0,T_M]$ in Eq.\ (\ref{eq:E}), we define $E_k(\bar{x})$ (without explicit $T$ notation) as
\begin{equation}\label{eq:Ek}
E_k^2(\bar{x})\equiv\frac{1}{p}\sum_{j=1}^p E_k^2(T_j,\bar{x})
\end{equation}
where the integral has been replaced for a sum over the actual $p$ sampled times $T_j$ in $[0,T_M]$.

We estimate the size of the neighborhood as
\begin{equation}\label{eq:eps_k}
\epsilon_k^2(\bar{x})\equiv\frac{2}{k(k+1)}\sum_{\substack{\bar{x}',\bar{x}''\in\mathcal{U}_k(\bar{x})\\ \bar{x}''\neq \bar{x}'}}|\bar{x}'-\bar{x}''|^2,
\end{equation}
which is a robust measure of the characteristic square radius of $\mathcal{U}_k(\bar{x})$. Notice that here we are making no assumptions on the box counting dimension of this set.

Finally, the noise amplification estimated from $k$ nearest neighbors is \cite{comment:resolution}
\begin{equation}\label{eq:sigma_k}
\sigma_k^2(\bar{x})\equiv\frac{E_k^2(\bar{x})}{\epsilon_k^2(\bar{x})},
\end{equation}
and we consider the average over $N$ reference points $\bar{x}_i$ of the reconstructed attractor randomly selected from the time series to define the global measure
\begin{equation}\label{eq:meansigma_k}
\sigma_k^2\equiv\frac{1}{N}\sum_{i=1}^{N}\sigma_k^2(\bar{x}_i).
\end{equation}
For this $k$-NN estimation of $\langle\sigma^2\rangle$ a new free parameter has been introduced to the method, namely the number of neighbors $k$. If the value of $k$ is chosen too large, the linear approximation of $g^T$ around $\bar{x}$ will be invalid and $\sigma_k^2(\bar{x})$ will differ from the $\epsilon\rightarrow 0$ limit $\sigma^2(\bar{x})$. On the other hand, $k$ needs to be large enough for the convergence of the estimator of the conditional variance (Eq.\ \ref{eq:Vark}). In Section \ref{sec:fnn} we will present practical considerations concerning the selection of appropriate values for $k$.

\subsection{Normalization}\label{sec:norm}

In the framework of the new interpretation of $\sigma^2(\bar{x})$ in terms of $\|\nabla g^T(\bar{x})\|^2$ discussed in Section \ref{sec:complexity}, the value of $\epsilon$ is no longer related to the observational noise and therefore its scale can change with the reconstruction.
For example, were $y=\varPsi(\bar{x})$ the optimal reconstruction, a simple rescaling $\alpha y$ would also be optimal because it would leave the neighborhood structure unchanged and therefore would not affect the computation of any dynamical invariant nor the application of any prediction algorithm.
However, in the calculation of $\sigma_k$ the characteristic radii $\epsilon_k$ would be affected by the factor $\alpha$, and therefore the resulting $\sigma_k$ will also differ by a factor $\alpha$. This is clearly undesirable: equally optimal reconstructions should yield the same $\sigma_k$ values.

The hypothetical case described above is somewhat artificial in the sense that it cannot occur when reconstructing a dynamical system: no such global scaling factor $\alpha$ will unexpectedly affect interpoint distances for a particular reconstruction among the set of all possible DC reconstructions which is obtained by varying parameters $\tau$ and $m$. However, a subtle effect will be present upon variation of these parameters: the value of $\epsilon_k(\bar{x})$ along the attractor will grow with larger delays due to the irrelevance effect which stretches (and folds) the attractor.

\begin{figure}
\centering
\includegraphics[width=0.48\textwidth]{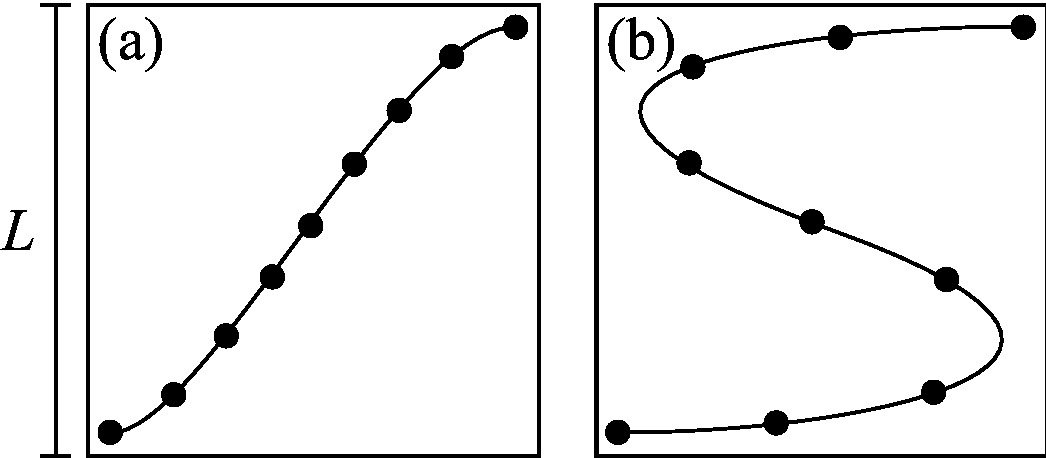}
\caption{Schematic representation of two delay coordinate (DC) reconstructions, (a) $\varPhi_1$ and (b) $\varPhi_2$, with $\tau_1<\tau_2$. The distance between neighbors increases with the delay time $\tau$ due to the stretching and folding of the attractor, while the global characteristic size $L$ of the attractor remains constant because it only depends on the amplitude of the time series.}
\label{fig:escalas}
\end{figure}

Figure \ref{fig:escalas} illustrates this behavior schematically.
Each panel depicts a DC reconstruction given by $\varPhi_1$ and $\varPhi_2$, where $\tau_1<\tau_2$.
The attractor, represented by a curve, is sampled the same number of times in both cases.
In this example the larger value of $\tau_2$ induces a stretching and folding of the reconstructed attractor.
However, the characteristic size $L$ of the reconstructed attractors is the same as it is determined by the amplitude of the time series.
On the other hand, the typical first neighbor distance is larger for $\varPhi_2$, an effect that is captured by larger values of $\langle\epsilon_k\rangle$.
However, larger values of $\langle\epsilon_k\rangle$ produce in turn a smaller value of $\sigma_k$, which goes in an opposite direction to a desired penalization of irrelevance.

In Fig.\ \ref{fig:epsvstau} we use the time series from the $x$ variable of the Lorenz system (see Appendix \ref{ap:lorenz} for details) to profile the behavior of $\epsilon_k$ defined by
\begin{equation}\label{eq:meaneps_k}
\epsilon_k^2\equiv\frac{1}{N}\sum_{i=1}^{N}\epsilon_k^2(\bar{x}_i),
\end{equation}
as a function of the reconstruction parameters $m$ and $t_w$.
\begin{figure}
\centering
\includegraphics[width=0.48\textwidth]{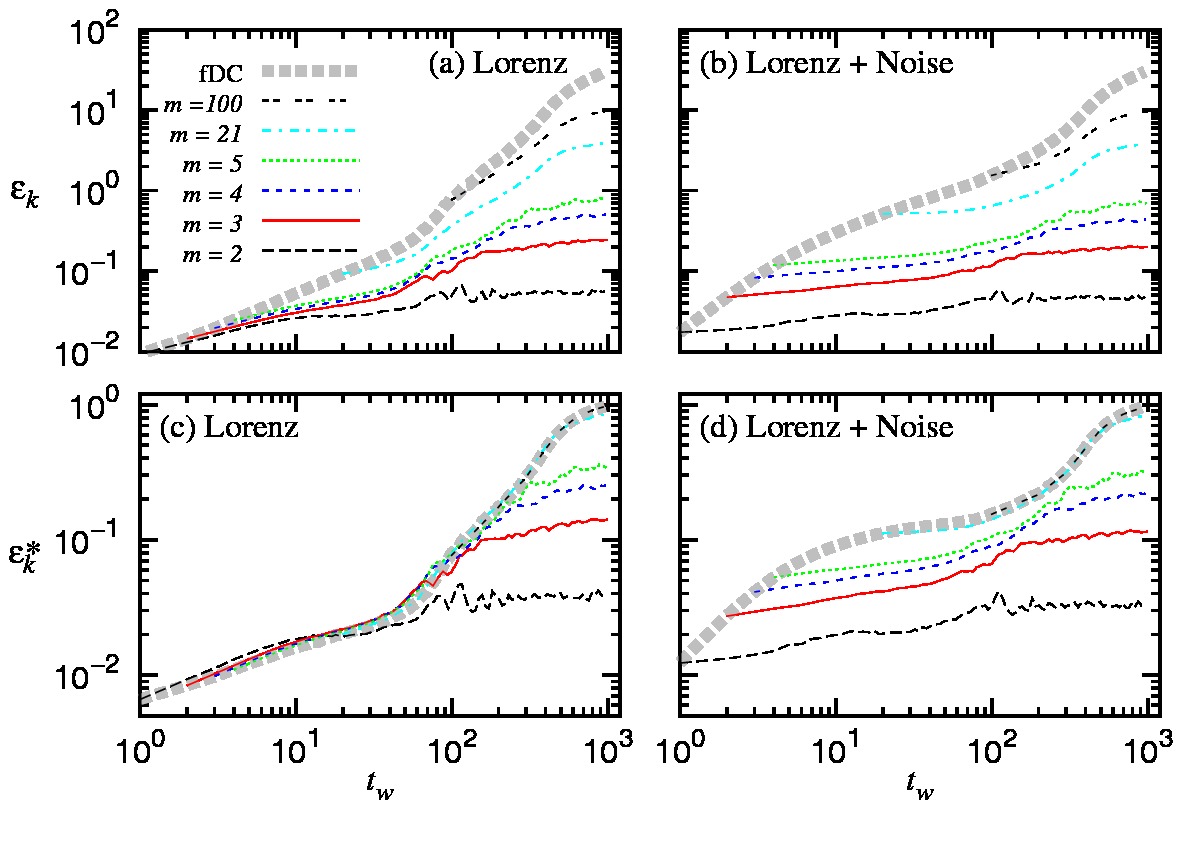}
\caption{(Color online). Response of the characteristic radii $\epsilon_k$ ($k=2$) to changes in the reconstruction parameters $t_w$ and $m$ for the Lorenz time series (noise free and with 10\% noise). (a) $\epsilon_k$ monotonically grows with $t_w$ and $m$ in the noise free case. (b) The same holds true for the noisy case. (c) For the normalized radii $\epsilon_k^*=\epsilon_k/\sqrt{m}$ the dependence with $m$ is eliminated at the lower end of $t_w$ values. (d) In the noisy case \---as well as for large $t_w$ in the noise-free case\--- the sampled points fill the $m$-dimensional space and the growth of $\epsilon_k$ with $m$ cannot be avoided. In all cases the thick dashed line corresponds to the full delay coordinate (fDC) reconstruction ($\tau=1$ and $m=t_w+1$).}
\label{fig:epsvstau}
\end{figure}
As argued above, this figure quantifies how the characteristic distance between first neighbors averaged over the attractor grows with the size of the considered time window.
Additionally, the typical distance between neighbors will also grow with the dimension induced by the noise which populates all directions as pointed out in \cite{Kennel:2002}.
We therefore conclude that in order to be able to compare different reconstructions a normalization is needed to account for this changing average interpoint distance.
We also notice that $\epsilon_k$ captures the local scale variations between reconstructions and is a measure of the degree of irrelevance of the considered delayed components.
Taking this argument further, the normalization we propose in this section follows from considering the average of $\sigma_k^2(\bar{x})$ over the attractor (Eq.\ \ref{eq:meansigma_k}) as a weighted average of $E_k^2(\bar{x})$ with weights $w_k(\bar{x})$ proportional to $\epsilon_k(\bar{x})^{-2}$, i.e.
\begin{equation}
w_k(\bar{x})=\alpha_k{\epsilon_k(\bar{x})^{-2}}.
\end{equation}
Across different reconstructions
these weights should satisfy $\sum_i w_k(\bar{x_i})=1$, therefore the normalization factor is
\begin{equation}\label{eq:norm2}
\alpha_k^2=\left[\sum_i \epsilon_k^{-2}(\bar{x}_i)\right]^{-1}
\end{equation}
in such a way that $\alpha_k^2\sigma_k^2=\sum_i w_k(\bar{x}_i)E_k^2(\bar{x}_i)$.

This normalization gives the product $\alpha_k\sigma_k$ the units of $x$, the observed variable, independently of the proposed reconstruction. This will allow a direct comparison between any kind of reconstruction for a given time series. If the statistics is so normalized, $\alpha_k\sigma_k$ will be upper bounded by the standard deviation of the time series and lower bounded by the noise level.

\subsection{Overview}

Here we collect the arguments of the previous sections to construct a cost function $L$ to guide the search of the optimal reconstruction.
Ideally, such function can be thought of as a sum of two terms with competing behavior
\begin{equation}\label{eq:L1}
L=R+\lambda I.
\end{equation}
The $R$ term should penalize redundance when the window size $t_w$ is too small or, in case it is not,
when the number of components within $t_w$ is insufficient to unfold the attractor.
On the other hand, the $I$ term should penalize irrelevance when the window size $t_w$ is too large or, in case it is not,
when the number of components within $t_w$ is unnecessarily larger than needed to unfold the attractor.

Throughout this section we have shown the basis and definition of $\sigma_k$ inspired on the definition of noise amplification in \cite{Casdagli:1991}, from which we arrived at a new interpretation in terms of the complexity of the dynamics $g^T$.
Furthermore, we have defined a normalization factor $\alpha_k$ needed to adjust $\sigma_k$ to scale variations induced by stretching and folding of the attractor when irrelevant delayed components enter the reconstruction vector.
These two terms, $\sigma_k$ and $\alpha_k$, are natural candidates to play the role of $R$ and $I$. We therefore define
\begin{eqnarray}
R_k&=&\log_{10}\sigma_k\\
I_k&=&\log_{10}\alpha_k
\end{eqnarray}
and arrive to
\begin{eqnarray}
L_k&=&R_k+I_k\\
&=&\log_{10}(\alpha_k\sigma_k)
\end{eqnarray}
where the parameter $\lambda$ in Eq.\ (\ref{eq:L1}) must necessarily be equal to 1, following the conception of $\alpha_k$ as a normalization factor (Section \ref{sec:norm}).

In Appendix \ref{ap:algorithm} we give details on our implementation of the method proposed in this work, which is freely available.

\section{Related work}\label{sec:related}

\subsection{Selection of $k$ and relationship with the method of FNN of Kennel {\it et al}.}\label{sec:fnn}

If we only consider the first neighbor to compute $\sigma_k(\bar{x})$, i.e.\ $k=1$, and choose the time step $T$ equal to $\tau$ reducing the sum in Eq.\ (\ref{eq:Ek}) to a single term, we exactly recover the statistics proposed by Kennel {\it et al}.\ \cite{Kennel:1992} to detect false nearest neighbors, that is
\begin{equation}\label{eq:equivKennel}
\sigma_1(T=\tau,\bar{x})=\frac{|x(t+\tau)-x'(t+\tau)|}{d(\bar{x},\bar{x}')}
\end{equation}
where $\bar{x}'$ is the first neighbor of $\bar{x}$ in the reconstruction and $d(\bar{x},\bar{x}')$ their distance and we question whether it is a true neighbor.
The criterion used by Kennel {\it et al}.\ was to consider $\bar{x}$ and $\bar{x}'$ false neighbors if $\sigma_1(T=\tau,\bar{x})>10$.

\begin{figure}
\centering
\includegraphics[width=0.48\textwidth]{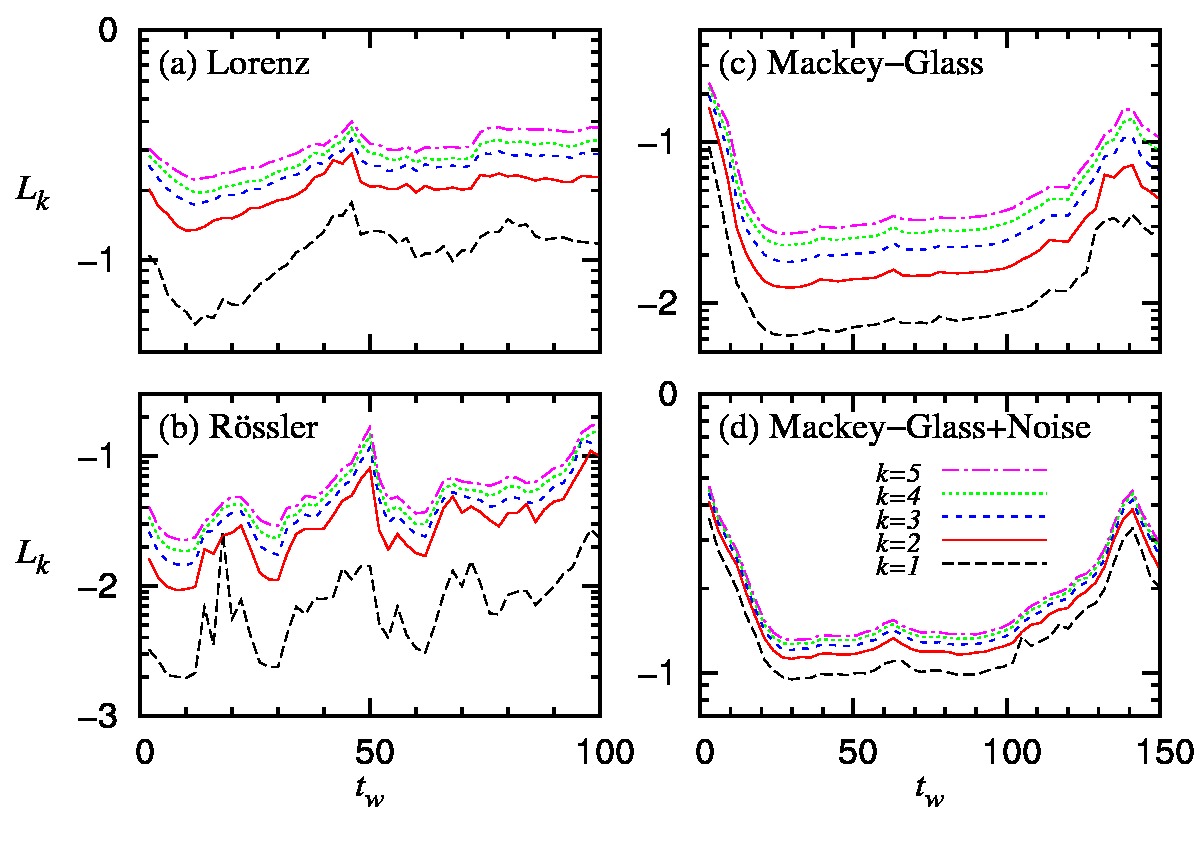}
\caption{(Color online). Profiles of $L_k$ {\it vs.}\ $t_w$ for a range of $k$ values and benchmark case studies. (a) Lorenz, $m=3$. (b) R\"ossler, $m=3$. (c) Mackey-Glass, $m=4$. (d) Mackey-Glass with 10\% noise, $m=4$. For all time series studied in this work the structure of maxima and minima converges for low values of $k$.}
\label{fig:k}
\end{figure}

The method of Kennel {\it et al}.\ is therefore encompassed by our proposal as a particular case with $k=1$.  The main difference is given by the fact that we consider $T$ in the interval $[0,T_M]$ instead of $T=\tau$. The time lag parameter $\tau$, or equivalently $t_w$, must also be determined.
Figure \ref{fig:k} shows the profiles of $L_k$ {\it vs.}\ $t_w$ for several values of $k$ and benchmark time series.
Choosing $k=1$ is not always the best option because the obtained profiles can be too noisy for a correct determination of the optimal $t_w$.
Increasing the value of $k$ regularizes the estimation of $Var(x(T)|B_\epsilon(\bar{x}))$ but also increases the size of $\mathcal{U}_k(\bar{x})$.
The corresponding profiles of $L_k$ tend to be smoother or converge to a stable profile but the method becomes less sensitive to local divergences.
A trade-off solution is given by the smallest value of $k$ for which the $L_k$ profile is stable in the sense of consistency in the optimal $t_w$ values obtained, which are given by the position of the global minima of $L_k$.
In general, we found $k=2$ or $3$ to be a good choice for the time series considered in this paper, and suggest the use of these values for general applications.

\subsection{Relationship with prediction error}\label{sec:prediction}

Here we analyze the mean of $E_k^2(T,\bar{x})$ (from Eq.\ (\ref{eq:Vark})) over the attractor instead of the proposed cost function.
This is nothing else than the usual mean squared prediction error (MSE), which we will here compute using a local constant model based on the first $k$ neighbors and will note $E_k^2(T)$.

\begin{figure}
\centering
\includegraphics[width=0.48\textwidth]{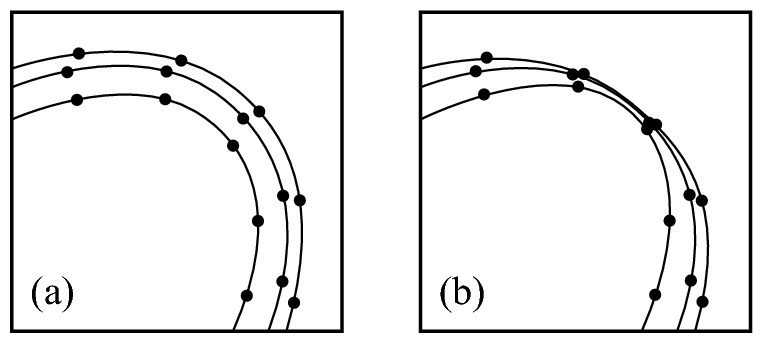}
\caption{Schematic representation of two reconstructions of the same region of an attractor with (a) well behaved orbits and (b) collapsed ones. The computation of $E_k(T)$ using $k=2$ will yield identical results for both reconstructions while $L_k$ will penalize the collapsed orbits of panel (b).}
\label{fig:orbits}
\end{figure}

\begin{figure}
\centering
\includegraphics[width=0.48\textwidth]{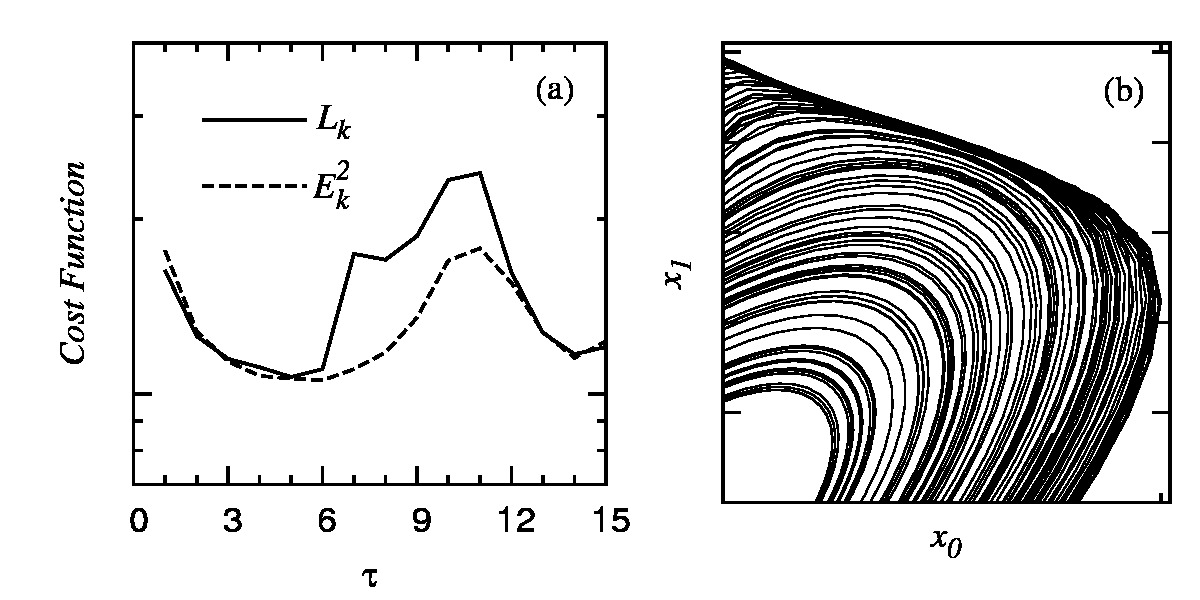}
\caption{Illustration of the sensitivity of $L_k$ and $E_k$ to an orbit collapse for the 3-dimensional delay coordinate (DC) reconstruction of the R\"ossler system. (a) Plot of $L_k$ and $E_k$ vs.\ $\tau$.  Notice that at $\tau=7$ the curves differ maximally. (b) Detail of the reconstructed attractor at $\tau=7$. The reconstruction exhibits collapsed orbits at this specific delay value.}
\label{fig:Pred}
\end{figure}

The question we address in this Section is whether a minimization of $E_k(T)$ constitutes an appropriate criterion to select reconstruction parameter values instead of the more complex definition of $L_k(T)$.
The main difference between these two approaches is that $E_k(T,\bar{x})$ does not carry any information about the size of the neighborhood $\mathcal{U}_\epsilon(\bar{x})$.
The relevance of this lack of information becomes evident when considering a reconstruction with a region of collapsed orbits.
By `collapsed orbits' we here refer to the case of true neighbors that become arbitrarily close for a given reconstruction ---without involving the presence of false neighbors.   This is illustrated in Fig.\ \ref{fig:orbits} and again on panel (b) of Fig.\ \ref{fig:Pred}.
Panels (a) and (b) show two different reconstructions of the same group of points; as we can see in panel (b) the orbits are collapsed.
If we compute $E_k(T)$ using $k=2$ we arrive at identical values for both reconstructions.  This follows from the fact that the distortion occurring in panel (b) does not alter the neighborhood structure.
On the other hand, if we compute $\sigma_k(T,\bar{x})$ for both reconstructions we will arrive at different values.
The case of panel (b) will yield a larger value of $\sigma_k(T,\bar{x})$.  This will follow from smaller neighborhood sizes in the denominator for regions of the reconstruction that exhibit a collapse.

The situation of an orbit collapse described in this Section is typical of the `redundance limit' (small $t_w$), but can also occur for intermediate $t_w$ values in low dimensional reconstructions.
This effect can be illustrated for the three dimensional DC reconstruction of the R\"ossler system from the time series of the $x$ variable (see Appendix \ref{ap:rossler}).
For $\tau=7$ there is an orbit collapse as shown in Fig.\ \ref{fig:Pred}(b).
Figure \ref{fig:Pred}(a) shows the response of $E_k$ as compared to $L_k$ for this three dimensional reconstruction as a function of $\tau$.
Notice how the profile of $L_k$ jumps at $\tau=7$ while $E_k$ does not.

\begin{figure}
\centering
\includegraphics[width=0.48\textwidth]{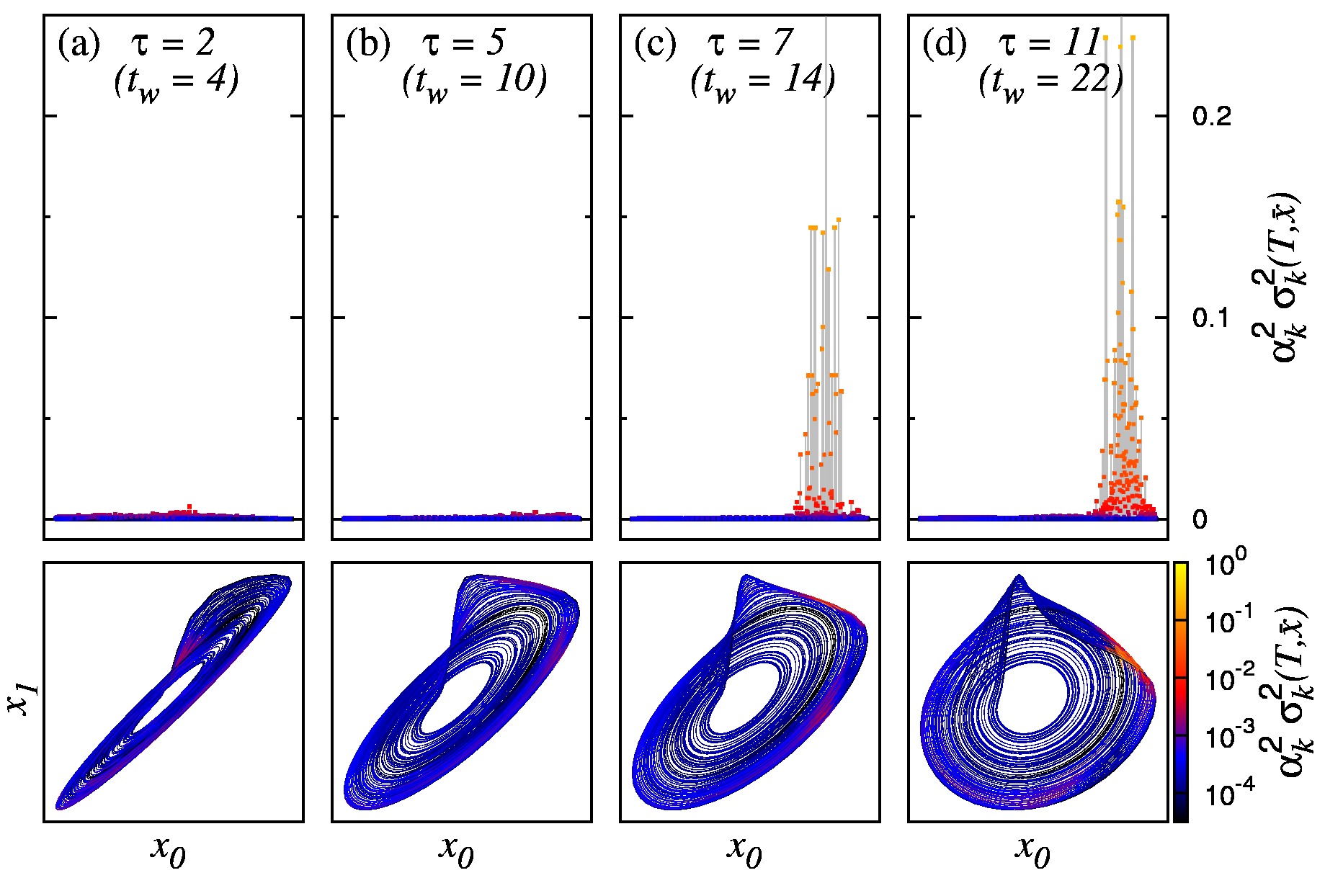}
\caption{(Color online). Sensitivity of $L_k$ to collapsed orbits. The local cost function $\alpha_k^2\sigma_k^2(T,\bar{x})$ is plotted vs.\ $x_0$ in the upper panels and color-coded in the two dimensional projection of the reconstruction in the lower panels for reconstruction parameters $m=3$ and (a) $\tau=2$, (b) $\tau=5$, (c) $\tau=7$ and (d) $\tau=11$.}
\label{fig:colorSigma}
\end{figure}

To support the contention that the response of $L_k$ shown in Fig.\ \ref{fig:Pred}(a) corresponds to the collapsed region of Fig.\ \ref{fig:Pred}(b) we use $\alpha_k^2\sigma_k^2(T,\bar{x})$ as a local cost function to extract spatial information about the reconstruction quality.
This has been done in Fig.\ \ref{fig:colorSigma} where the values of $\alpha_k^2\sigma_k^2(T,\bar{x})$ have been plotted on the attractor in the lower panels and as a function of $x(t)$ ($=x_0$) in the upper panel.
From left to right the reconstructions have $\tau=2$, $\tau=5$, $\tau=7$ and $\tau=11$ respectively.
For $\tau=7$ (panel (c)) the values of $L_k$ are clearly dominated by the behavior in the collapsed region.
The same holds true for $\tau=11$ (panel (d)) but in this case there are false neighbors as orbits cross each other \cite{comment:Fraser}.

\begin{figure}
\centering
\includegraphics[width=0.48\textwidth]{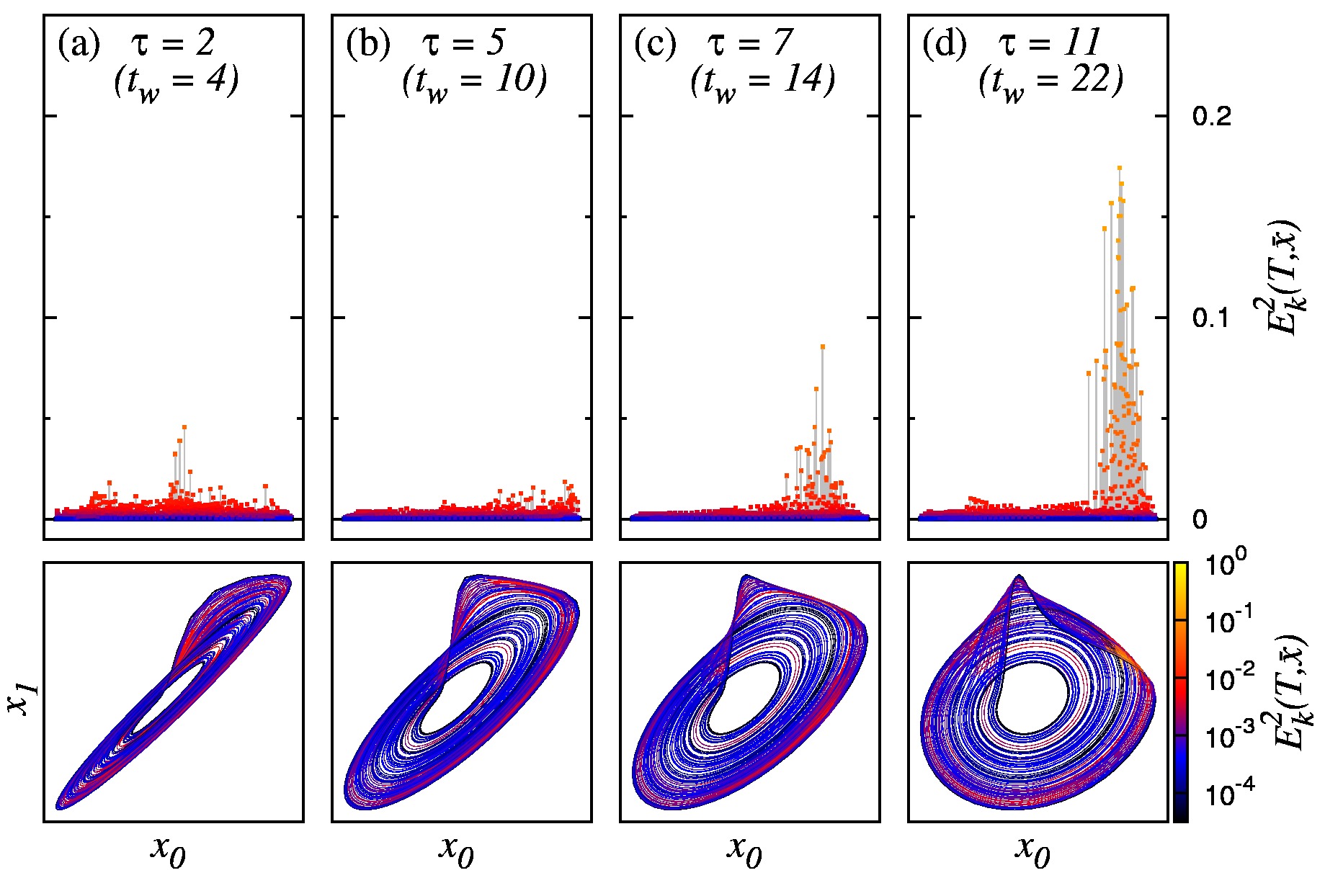}
\caption{(Color online). Same as Fig.\ \ref{fig:colorSigma} for $E_k^2(T,\bar{x})$. This quantity is less sensitive to a collapse of orbits (panel (c)) and false neighbors (panel (d)) as compared to its ground level, which presents a spurious structure (panels (a) and (b)). See the main text for details.}
\label{fig:colorVarx}
\end{figure}

If we compare Fig.\ \ref{fig:colorSigma} with \ref{fig:colorVarx}, where $\alpha_k^2\sigma_k^2(T,\bar{x})$ has been replaced by $E_k^2(T,\bar{x})$, we find that the local response to the collapse of orbits ($\tau=7$) and to the presence of false neighbors ($\tau=11$) is lower, especially as compared to the ground response.
We note in passing that this ground response is in itself an interesting finding from this plot.  More precisely,
$E_k^2(T,\bar{x})$ exhibits a rich structure along the attractor (e.g.\ along the radial direction on the spiral) even when the reconstruction is optimal ($\tau=5$, panel (b)).
$E_k^2(T,\bar{x})$ penalizes orbits of the attractor that do not seem to be more divergent than others where however $E_k^2(T,\bar{x})$ is smaller.
Instead, these orbits belong to lower density regions, which yields larger values of $E_k^2(T,\bar{x})$.
As expected, $E_k^2(T,\bar{x})$ evaluates the quality of the prediction algorithm (k-NN in this case) rather than reconstruction quality.
On the other hand, the variability of $\alpha_k^2\sigma_k^2(T,\bar{x})$ across the attractor shown in the lower panel of Fig.\ \ref{fig:colorSigma}(b) is less severe and only associated to intrinsic divergence of orbits.

In Small {\it et al.}\ \cite{Small:2004} the authors explored the use of the $k$-NN prediction algorithm for the selection of optimal reconstruction parameters.
The quantity they proposed to minimize is the {\it description length} (DL) of the time series using this particular modeling approach and a candidate reconstruction.
In the particular case of $k$-NN the description length reduces to the logarithm of $E_k(T)$; therefore, the previous critiques in this Section apply to their method.
Another important drawback is that their analysis of the method is reduced to $k=1$ and $T=1$.
The authors also referred to previous works based on minimizing DL for different prediction algorithms (radial basis functions \cite{Judd:1995} and neural networks \cite{Small:2002}).
They concluded that for these algorithms DL is harder to estimate, and that their free parameters are harder to optimize for the purpose of determining reconstruction parameters.

\subsection{Relationship with dynamical methods}\label{sec:related_dynamical}

The dynamical methods \cite{Buzug:1992a, Liebert:1991, Gao:1993, Kennel:2002} mentioned in the Introduction are also based on the detection of FNNs and provide a criterion for choosing $m$ and $\tau$.
The methodology we propose in this work can also be considered a dynamical method.
In the following we will discuss the most relevant differences between these methods and our approach.

\subsubsection{Gao and Zheng}\label{sec:Gao}
In \cite{Gao:1993} Gao and Zheng used a quantity close to $\sigma_1(T,\bar{x})$ (notice $k=1$) to determine the reconstruction parameters $m$ and $\tau$, and also to estimate the maximum Lyapunov exponent of the dynamics.
This quantity is given by
\begin{equation}\label{eq:Gao}
R(T,\bar{x})=\frac{d(\bar{x}(T),\bar{x}'(T))}{d(\bar{x},\bar{x}')}
\end{equation}
where the only difference with $\sigma_1(T,\bar{x})$ is that both numerator and denominator are given by distances computed in the reconstructed space.
They then defined the cost functions to be minimized as $\Lambda= \langle ln(R(T,\bar{x}))\rangle$ and $\Lambda_+= \langle ln(R(T,\bar{x}))\rangle_+$, where in $\Lambda_+$ the mean is taken over positive values of $ln(R(T,\bar{x}))$ only.
Mean values of logarithmic divergence are suboptimal for the purpose of detecting collapsed orbits. %, as discussed in Section \ref{sec:Lyap}.
However, the main critique we make to the method of Gao and Zheng is the use of a distance in the full reconstructed space in the numerator to measure divergence of neighboring orbits.
Given a DC vector $\bar{x} = (x(t-(m-1)\tau),... x(t-\tau), x(t))$ at time $t$, its image under the dynamics for a time step $T$, $\bar{x}(T)$, will share $m-1$ components with $\bar{x}$ when $T=\tau$.
This constraint, which is inherent to the DC reconstruction and independent of the time series under analysis, will in general affect the distance $d(\bar{x}(T),\bar{x}'(T))$ in the numerator of Eq.\ (\ref{eq:Gao}) producing an artificial minimum at $\tau=T$ in a $\Lambda$ {\it vs.}\ $\tau$ profile.
\begin{figure}[h]
\centering
\includegraphics[width=0.48\textwidth]{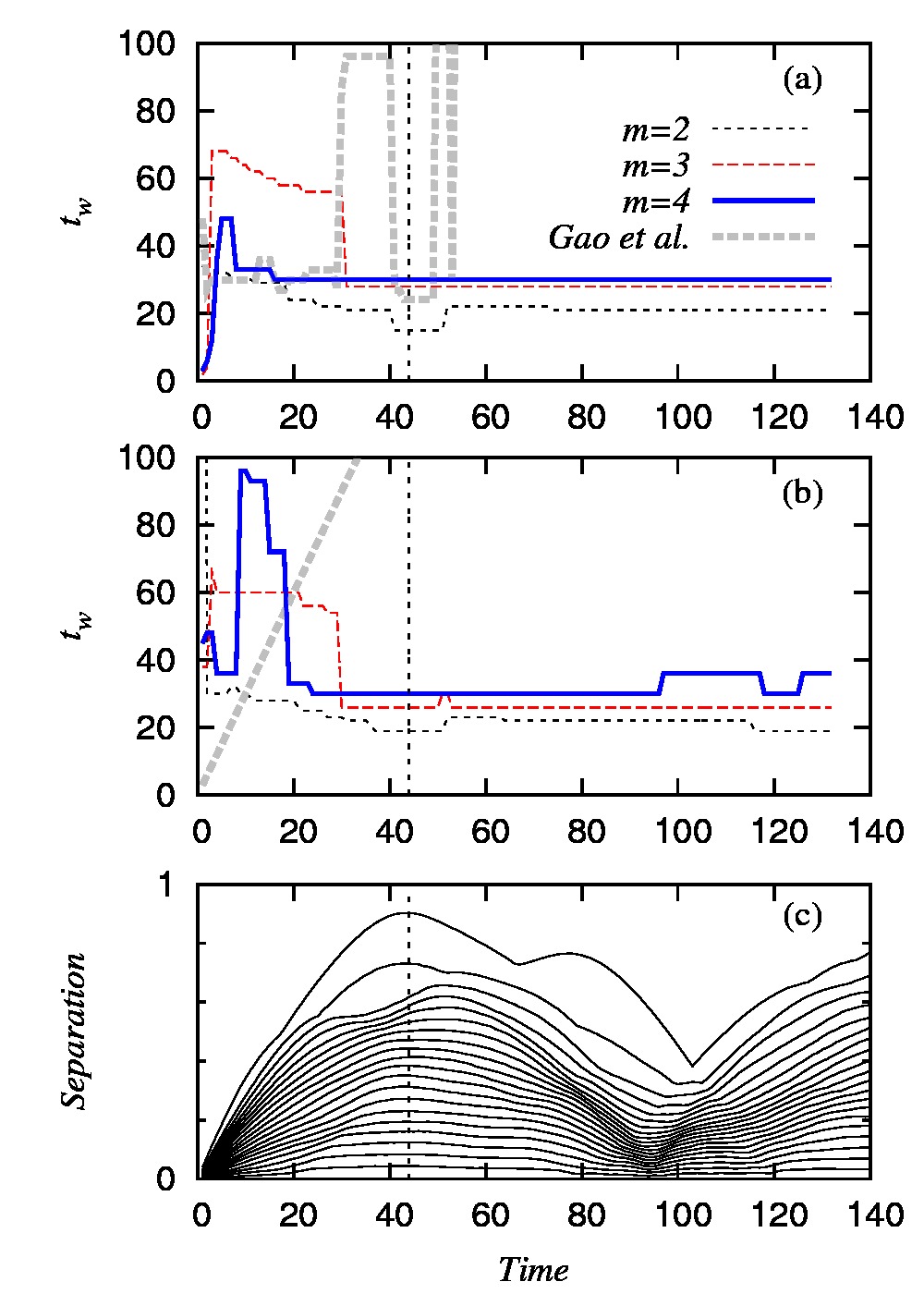}
\caption{(Color online). Optimal window size $t_w$ (in an $L_k$ sense) as a function of the horizon parameter $T_M$ for (a) the noise-free Mackey-Glass case and (b) the noisy case.  Panels (a) and (b) also show $t_w$ vs.\ $T$ as obtained by applying Gao's method (dashed gray line) which exhibits a high dependence on parameter $T$.  The dashed vertical lines signal the horizon parameter $T_M$ equal to the first maximum of the upper curve of the space-time separation plot \cite{Provenzale:1992} for the time series without any reconstruction (panel (c)).}
\label{fig:vsTm}
\end{figure}

In Fig.\ \ref{fig:vsTm} we plot the time window $t_w$ which minimizes $L_k$ as a function of parameter $T_M$ for the DC reconstructions of the noise-free Mackey-Glass time series (panel (a)) and also for the noisy case (adding 10\% i.i.d.\ noise, panel (b)).
For increasing dimensions $m$ from 2 to 4 we see that the time window converges to a stable value if $T_M$ is large enough.
In contrast, for small values of $T_M$ the value of $t_w$ is unstable.
This is due to the high correlation between successive values of oversampled time series which implies that $x(T)$ will be determined by $\bar{x}$ independently of the reconstruction quality.
In order to characterize a proper interval $[0,T_M]$ for $T$ we built a {\it space-time separation plot} \cite{Provenzale:1992} for the time series under consideration without any reconstruction as shown in Fig.\ \ref{fig:vsTm}(c).
The upper curve corresponds to the 95\% percentile of the distribution of $|x(t+T)-x(t)|$, i.e.\ this curve is almost an upper bound for $|x(t+T)-x(t)|$.
We use this curve to choose $T_M$ as the time of its first maximum in order to ensure that nearby orbits are given enough time to diverge.
For this time series we obtain $T_M=44$, a value that avoids the fluctuations in $t_w$ shown in panels (a) and (b).
Notice however that any value of $T_M$ between 30 and 100 is equally valid, and that no particular choice of this parameter is critical in order to obtain consistent results.
The use of the 95\% percentile curve from the space-time separation plot has given appropriate $T_M$ values that avoid fluctuations in $t_w$ for all time series considered in this paper (plots of $t_w$ vs.\ $T_M$ not shown).

In \cite{Gao:1993} the Authors give no suggestion to choose an adequate value for $T$.
Indeed, no value of $T$ ensures an independent output for $\tau$ as discussed previously in this Section.
As shown in Fig.\ \ref{fig:vsTm}(a) the $t_w$ obtained with Gao's method has a strongly fluctuating dependency on $T$.
This situation even deteriorates when noisy time series are considered (panel (b)), in which case one identically obtains $\tau = T$ as previously explained.

\subsubsection{Buzug and Pfister}
Close to the definition of $\Lambda$ by Gao and Zheng is the {\it averaged local deformation} proposed by Buzug and Pfister in an earlier work \cite{Buzug:1992a}.
Their divergence measure is obtained similarly to $\Lambda$ but with a small difference in the expression of Eq.\ (\ref{eq:Gao}):
the distance from the reference point $\bar{x}$ to its first neighbor $\bar{x}'$ is replaced by the distance to the center of mass of the cloud of neighbors initially inside a fixed-radius-ball around the reference point (both in the numerator and denominator).
Buzug and Pfister give no justification for the latter choice, which has two drawbacks:
first, as recognized by the Authors, in noisy conditions the center of mass tends to be close to the reference point, introducing a divergence in Eq.\ (\ref{eq:Gao}).
Second, the divergence of false neighbors is underestimated. More precisely, if the cloud is split when the dynamics evolves (assuming that there were false neighbors), the distance of $\bar{x}$ to the center of mass is about half of the distance between the two clouds. This effect makes it harder to discriminate false from true neighbors.

\subsubsection{Kennel and Abarbanel}
In 2002 Kennel and Abarbanel \cite{Kennel:2002} presented an improved version of their false nearest neighbor method \cite{Kennel:1992}.
This new method, which mainly deals with the case of oversampled time series, also produces an estimate of the optimal delay time.
The strategy is based on considering nearest strands, which are sets of nearest neighbors which are close in time and characterize nearest orbits rather than nearest neighbors.
The divergence of nearest neighbors is then averaged over neighbors in a strand and compared to a threshold.
The divergence measure is similar to Eq.\ (\ref{eq:equivKennel}), also with $T=\tau$ but the denominator is absent (it is replaced by the standard deviation of the time series for normalization purposes).
The proposed methodology has two undesirable consequences.
First, the value of $T$ should be the same for all reconstructions independently of the candidate delay time $\tau$.
This leads to wrong conclusions about the optimal delay time, because
the smaller are $\tau$ and $T$, the lower is the expected divergence and therefore fewer false neighbors will be detected using a fixed threshold.
The authors correctly diagnosed this problem and proposed to apply a linear transformation to the reconstructed space before using the false strands algorithm.  The purpose is to spread out the attractor which, in the case of a small $\tau$, tends to be collapsed onto the identity line.
We found this solution unsatisfactory because it requires to transform the reconstruction we intend to evaluate.
Secondly, the suppression of the denominator in Eq.\ (\ref{eq:equivKennel}) prevents the detection of collapsed orbits as discussed in Section \ref{sec:prediction}.

\subsubsection{Summary}
In summary, common drawbacks of existing dynamical methods in the literature are the following:
i) Results in general depend on $T$ as in \cite{Gao:1993}.  Usually the value of $T$ is either fixed to be equal to $\tau$ as in \cite{Liebert:1991,Kennel:2002},
or equal to the sampling time as in \cite{Buzug:1992a}, which is suboptimal and sampling dependent.
ii) The global reconstruction quality measure is an average of the logarithm of a divergence metric which does not fully capture the presence of false neighbors as in \cite{Liebert:1991,Gao:1993,Buzug:1992a} or is a count of divergences above a threshold which is fixed {\it ad hoc} as in \cite{Kennel:2002}.
iii) The number of neighbors is fixed to $k=1$ as in \cite{Liebert:1991,Gao:1993,Kennel:2002,Small:2004} while in some cases it is desirable to consider $k>1$ to gain robustness.

However, the main contribution in our proposal is not how to deal with these drawbacks but to introduce the normalization factor $\alpha_k$ which penalizes irrelevance by adjusting the scales in the reconstructed attractor.
We have not found in the literature any irrelevance measure based on the characteristic distance among neighbors.
This measure explicitly uses the fact that irrelevance stretches (and folds) the attractor.
Furthermore, this irrelevance measure is incorporated into the cost function as a normalization factor avoiding the introduction of extra parameters.
The factor $\alpha_k$ is the key ingredient in our reconstruction quality measure ---it is responsible for an absolute minimum in our cost function when screening all possible values of $t_w$.

\section{Practical applications}\label{sec:application}

\subsection{Noisy time series}\label{sec:noisy}

\begin{figure}
\centering
\includegraphics[width=0.48\textwidth]{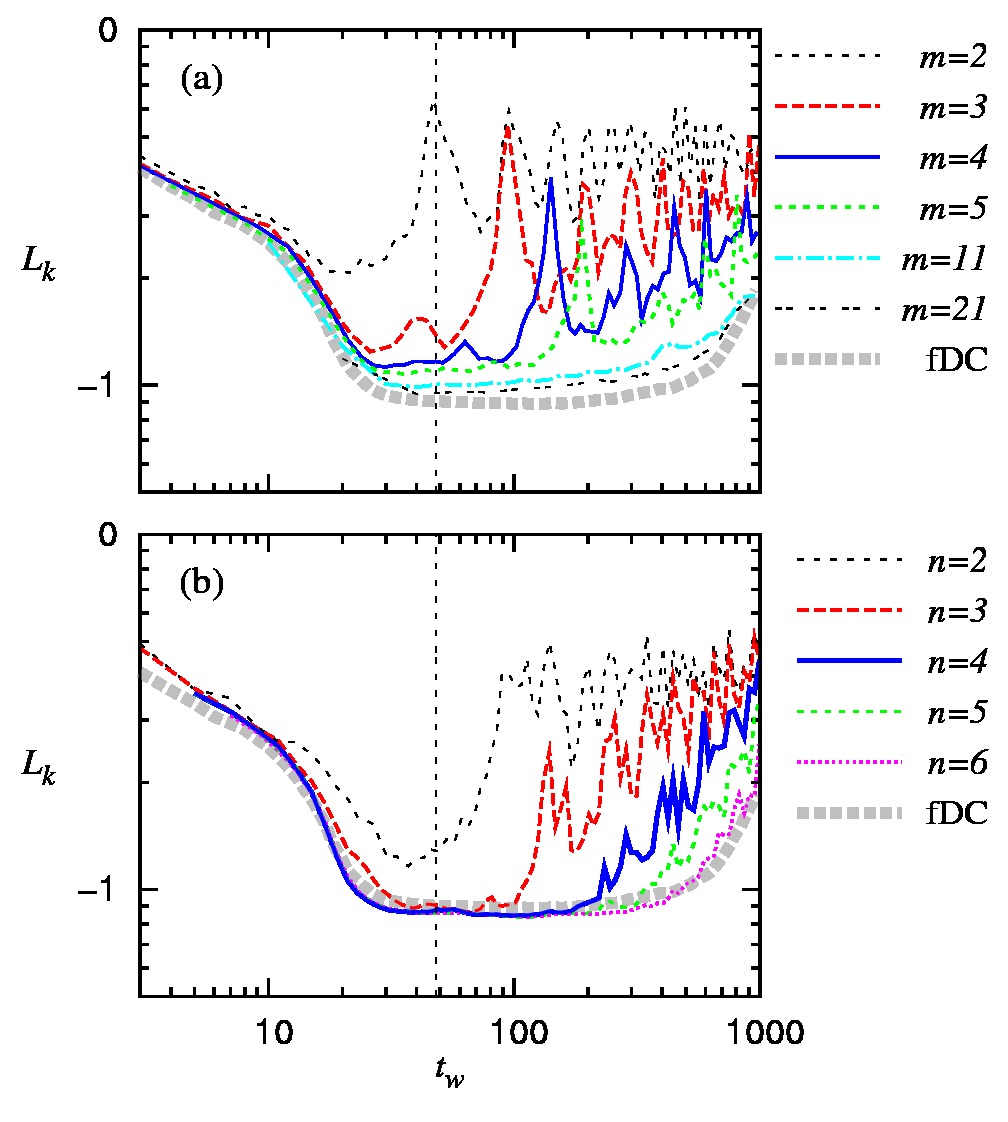}
\caption{(Color online). Cost function vs.\ time window size $t_w$ for the Mackey-Glass series plus 10\% noise. (a) Delay coordinate (DC) reconstructions with different dimensions $m$ as in Figure \ref{fig:example_mcgl}. (b) Legendre coordinates reconstruction of different dimensions $n$. Both panels show the curve for the fDC reconstruction ($\tau=1$ and $m=t_w+1$). Note that the Legendre profiles reach values below the latter curve. The vertical dashed line indicates $t_w=\tau^*_w$.}
\label{fig:noileg}
\end{figure}

For the case of noisy time series we found, in general, that the higher the number of delayed components considered inside the selected time window, the lower the value of $L_k$ is.
An illustration of this behavior is given in Fig.\ \ref{fig:noileg}(a) where we show the profiles of $L_k$ vs.\ $t_w$ for the Mackey-Glass time series with 10\% (amplitude) i.i.d.\ observational noise.
For this time series the minimal embedding dimension is $m=4$ in the noise-free case (see Fig.\ \ref{fig:example_mcgl}).
However, in the noisy case the profiles of $L_k$ vs.\ $t_w$ do not reach their minimum for a small dimension $m$, and the minimum of $L_k$ corresponds to the case of including every possible delayed value inside the time window of the reconstructed vector.
We previously called this high dimensional vector the {\it full delay coordinate} (fDC) vector for a given time window, and it has a dimension $m=t_w+1$ (where $t_w$ is expressed in sampling time units) which depends on the sampling.
The behavior of $L_k$ can be explained by noticing that the inclusion of redundant components allows noise filtering (given that the noise is i.i.d.).
Noise filtering is performed in the computation of the Euclidean distances between pairs of reconstructed vectors since the Euclidean distance is essentially an average of the quadratic distance for each component.
More precisely, for large $m$ the distance $d_{noisy}^2$ between two noisy DC vectors tends to be \cite{Hegger:2001}
\begin{equation}\label{eq:Hegger}
d_{noisy}^2\approx d_{clean}^2+2m\xi^2
\end{equation}
where $d_{clean}^2$ indicates distance between clean vectors and $\xi^2$ is the noise variance.
As the second term does not depend on the specific pair of vectors (assuming i.i.d.\ noise), $d_{noisy}^2$ produces a correct rank of nearest neighbors.
This result was anticipated in \cite{Casdagli:1991} from an information point of view: more information implies less distortion, therefore the distortion is a monotonic, nonincreasing function of the dimension.
Our cost function is reflecting this behavior because it is based on the theoretical definition of noise amplification $\sigma(T,\bar{x})$ as defined in \cite{Casdagli:1991}.

\subsection{Discrete Legendre polynomials}

In the previous sections we have only considered delay coordinate reconstructions, and found that high dimensional reconstructions yield minimum noise amplification in the case of noisy time series.
One can apply a further transformation $\varPsi:\field{R}^{m}\rightarrow\field{R}^n$ in order to reduce the dimensionality of the delay coordinate vector.
To this end we can use a linear $\varPsi$ which projects a fDC vector from $t_w+1$ dimensions onto $n$ directions given by, for example, the first $n$ discrete Legendre polynomials.
Therefore, the free parameters for the Legendre coordinates that we need to determine are the time window width $t_w$ and the dimension $n$.

Our aim is to show here how our cost function allows the evaluation of more general reconstructions than just delay coordinates.
Notice that the proposed method can be used to determine whether applying a further transformation $\varPsi$ improves the reconstruction quality.
In addition, it allows the computation of optimal parameters for this reconstruction exactly in the same way as for delay coordinate reconstructions.

Figure \ref{fig:noileg}(b) shows $L_k$ vs.\ $t_w$ for $n=1,\ldots,6$ for the the same noisy Mackey-Glass time series as in panel (a).
We see that the minimal embedding dimension $n=4$ is recovered despite the presence of noise.
However, the optimal time window width $t_w=113$ is much larger than in the noise-free case.
This difference can be explained by considering that, due to the presence of noise, the information about the system state carried by measures with larger delay times is now more relevant relative to the larger uncertainty on the system state (reflected by larger values of $L_k$ with respect to the noise free case).
As pointed out in Section \ref{sec:whenisoptimal} we expect that the optimal embedding parameters depend on the noise level.

Finally, we would like to notice that this transformation $\varPsi$ not only reduces the dimensionality from $m=t_w+1$ to $n=4$ but also the value of the cost function with respect to the lowest level attained with a DC reconstruction (which is obtained for the fDC vector and plotted in both panels of Fig.\ \ref{fig:noileg} with a gray dashed line).

\subsection{Chua's circuit}\label{sec:Chua}

\begin{figure}
\centering
\includegraphics[width=0.48\textwidth]{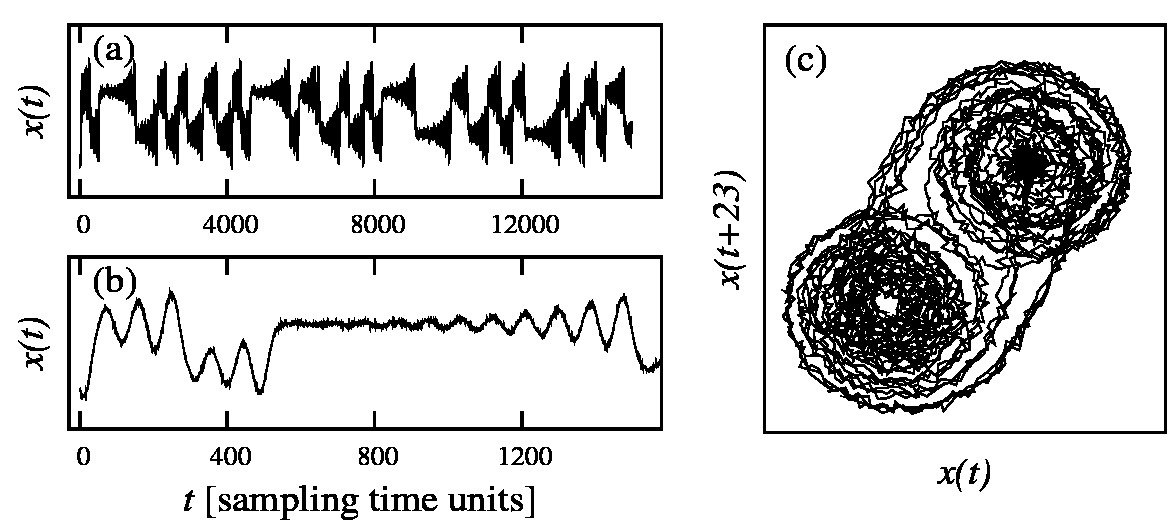}
\caption{Chua's circuit data \cite{Aguirre:1997}. (a) The full length time series. (b) Detail of the first 1500 data points. The characteristic period is approximately 94 sampling time units. (c) Preliminary 2-dimensional delay coordinate reconstruction using $\tau=23$ (which corresponds to a quarter of the characteristic period).}
\label{fig:chuadata}
\end{figure}

We now consider a real time series from a practical implementation of Chua's circuit.
The data correspond to measurements of inductor current values performed in \cite{Aguirre:1997} and can be retrieved from \cite{www:data1}.
The data are depicted in Fig.\ \ref{fig:chuadata}.
This time series has a significant amount of observational noise mainly due to a small resolution in the A/D conversion and to the Hall-effect probe used to measure the current through the inductor \cite{Aguirre:1997}.

\begin{figure}
\centering
\includegraphics[width=0.48\textwidth]{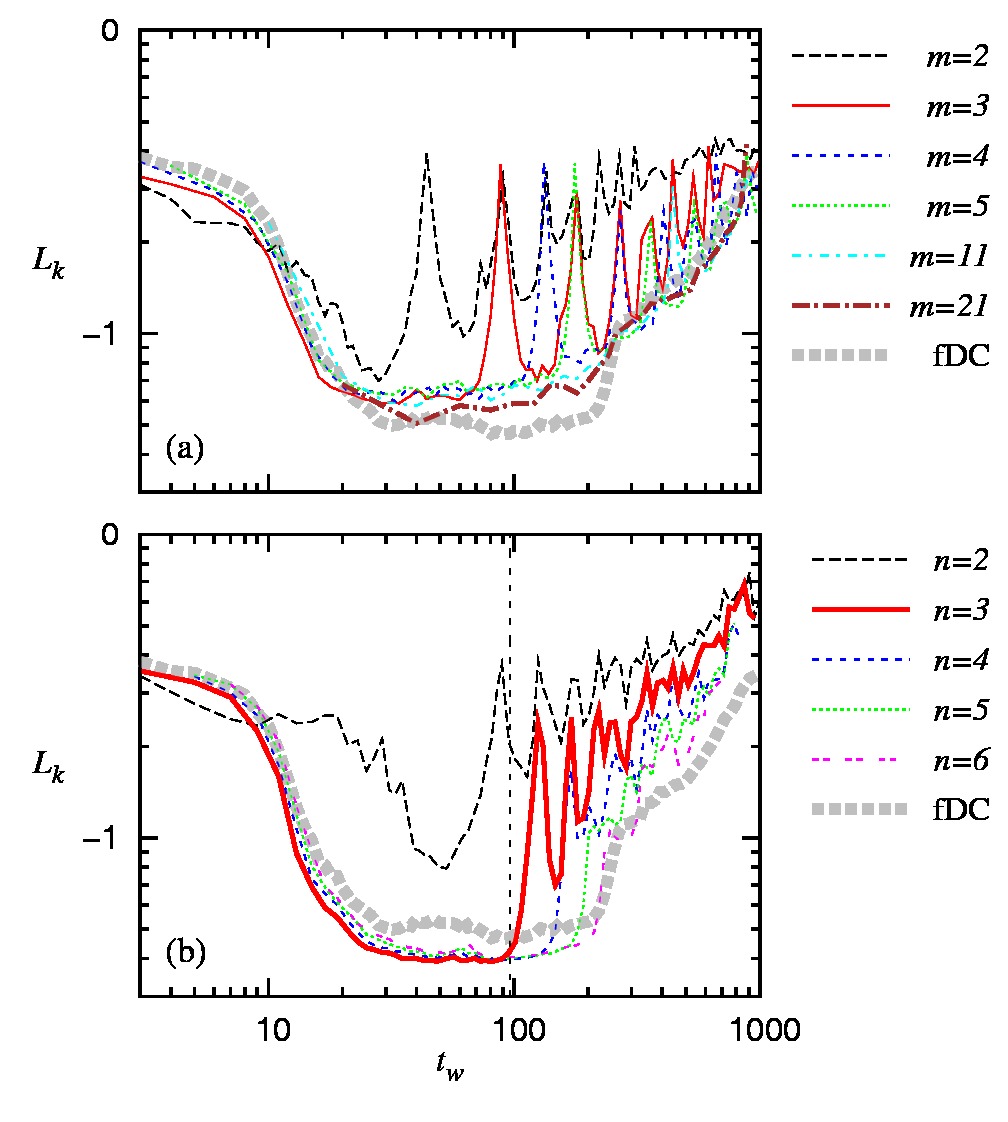}
\caption{(Color online). Cost function vs.\ time window size $t_w$ for Chua's circuit time series. Panel (a) corresponds to delay coordinate (DC) reconstructions with different dimensions $m$ as in Figure \ref{fig:example_mcgl}. Panel (b) corresponds to Legendre coordinates of different dimensions $n$. Both panels show the profile for the fDC reconstruction ($\tau=1$ and $m=t_w+1$). The vertical dotted line indicates the value of $\tau^*_w$.}
\label{fig:dsil}
\end{figure}

For this noisy time series we explored both delay and Legendre coordinates.
Figure \ref{fig:dsil}(a) shows the profile of $L_k$ vs.\ $t_w$ for DC reconstructions of increasing dimension $m$ from $2$ to $t_w+1$.
These results are consistent with the ones obtained for the noisy Mackey-Glass case (Fig.\ \ref{fig:noileg}(a)), where the minimum value of $L_k$ is reached when the DC vector incorporates all delayed values inside the selected time window.

In Fig.\ \ref{fig:dsil}(b) we show the curves $L_k$ vs.\ $t_w$ corresponding to Legendre coordinates for increasing dimensions $n$. For comparison we also plot the profile corresponding to the fDC reconstruction (plotted with a gray dashed line as in panel (a)).
Again, as observed for Legendre coordinate reconstructions of the noisy Mackey-Glass time series, the absolute minimum of $L_k$ is now reached for a low dimensional reconstruction with parameters $n=3$ and $t_w=81$.
This type of reconstruction significantly reduces the minimum value of $L_k$ obtained with a DC reconstruction, which is in turn due to the noise reduction effect of projecting onto the discrete Legendre polynomials.

Now we compute $\tau^*_w$ from Eq.\ (\ref{eq:tauCasdagli}) in order to find the upper bound for which the analytical solutions are equivalent to PCA.  Simply applying Eq.\ (\ref{eq:tauCasdagli}) to the raw data leads in general to an overestimation of $\langle (dx/dt)^2 \rangle$.
A solution is to smooth the data with a Savitzky-Golay filter \cite{Sav_Gol}.
This type of filter has two free parameters: the length of the fitting window and the order of the fitted polynomial.
Using a fitting window of 7 points and polynomials of order 2 we found $\tau^*_w\approx 96$.
The time window width obtained by minimizing $L_k$ is less than but close to the upper bound $\tau^*_w$, in agreement with the guidelines given in \cite{Gibson:1992} to choose $t_w$ in order to maximize the signal to noise ratio and simultaneously avoid irrelevance effects.
Furthermore, $t_w<\tau^*_w$ implies that the Legendre coordinate reconstruction which is optimal in terms of $L_k$ is likely performing a PCA over the fDC reconstruction.
\begin{figure}
\centering
\includegraphics[width=0.48\textwidth]{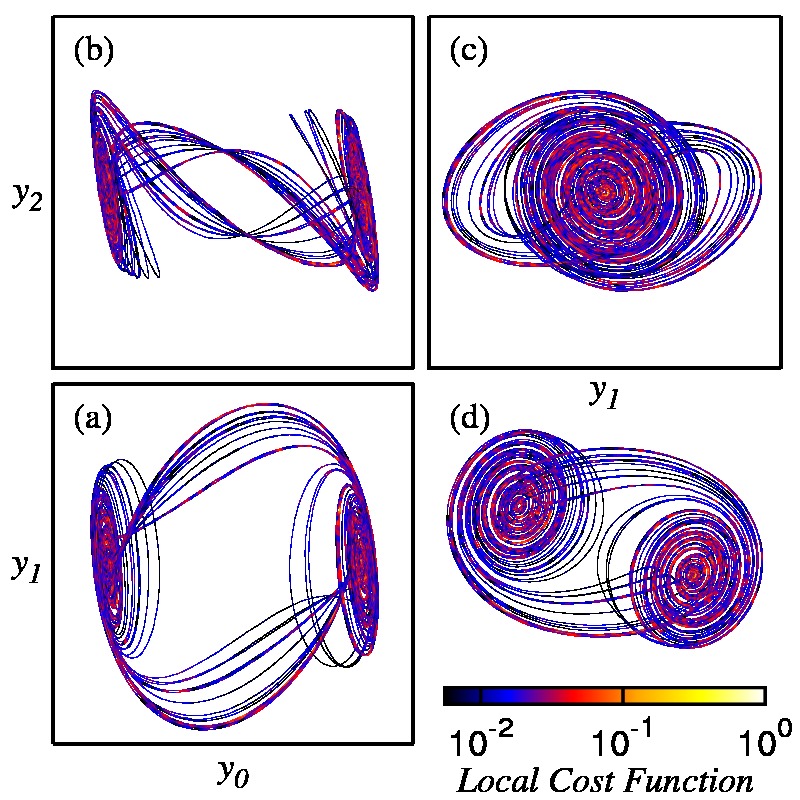}
\caption{(Color online). Views of the 3-dimensional Legendre coordinate reconstruction of Chua's circuit attractor. (a) $y_1$ vs.\ $y_0$, (b) $y_2$ vs.\ $y_0$ and (c) $y_2$ vs.\ $y_1$. (d) Perspective view of the reconstructed attractor. The local cost function over the attractor is color-coded. The principal directions of the attractor (in the PCA sense) are aligned with the coordinate axes. The local cost function does not present localized regions of high values (as e.g.\ in Fig.\ \ref{fig:color}(b)) implying the absence of false neighbors.}
\label{fig:color_dsil}
\end{figure}
This is confirmed by Fig.\ \ref{fig:color_dsil}, where from three independent views of the reconstruction we see that the Legendre coordinates are aligned with the principal directions of the attractor.

\subsection{SFI Laser}

\begin{figure}[h]
\centering
\includegraphics[width=0.48\textwidth]{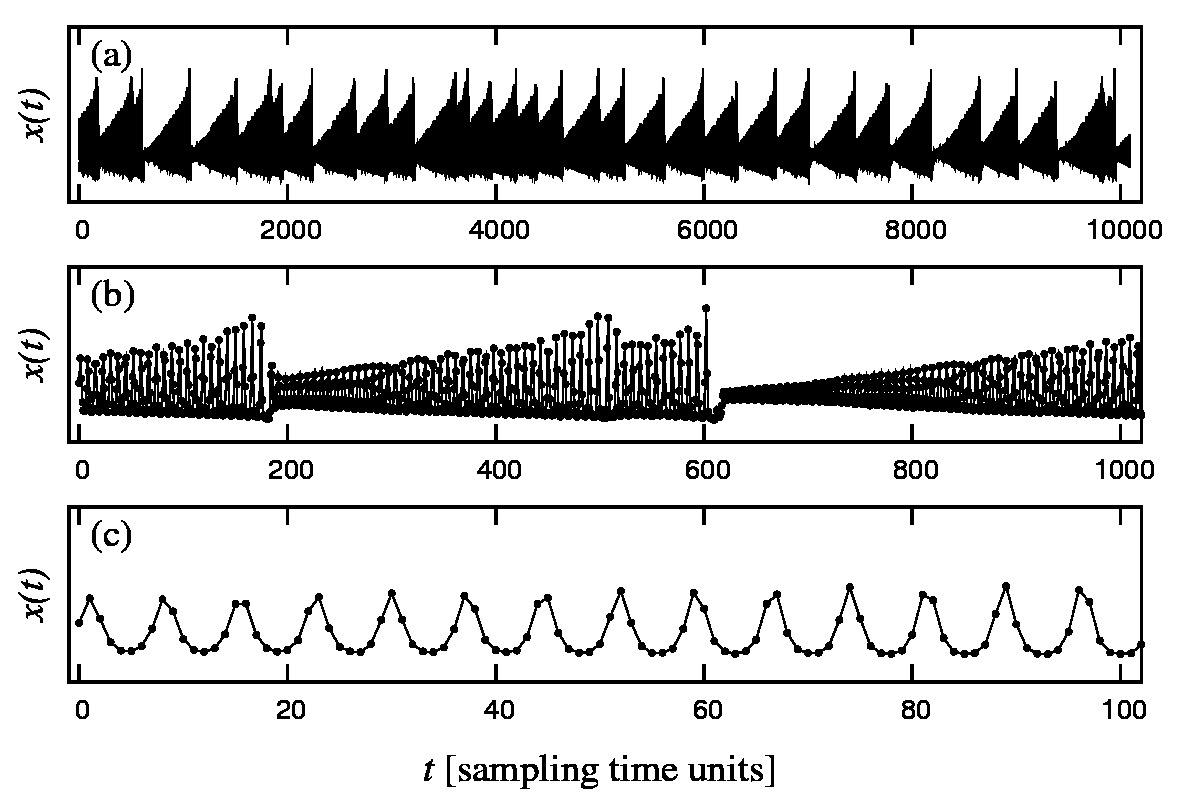}
\caption{SFI Laser data. (a) Full-length time series. (b) Detail of the first 1000 samples. (c) Zoom on the first 100 points. The characteristic period is approximately 7 sampling time units.}
\label{fig:laserdata}
\end{figure}

The last data set we consider in this work corresponds to measurements of fluctuations in a far-infrared laser taken from the Santa Fe Institute time series prediction competition \cite{Hubner:1994} and can be retrieved from \cite{www:data3}.
The full time series is plotted in Fig.\ \ref{fig:laserdata} panel (a), and further details on different time scales in panels (b) and (c).
The average pulsation frequency is approx.\ 1.65 MHz and the sampling time is 80 ns.
Therefore, the time series is sampled approximately 7.6 times per cycle (see panel (c) of Fig.\ \ref{fig:laserdata}), which is a very low sampling frequency as compared to the previous examples.
Notice that the time series covers a larger numbers of cycles than in previous examples.
This in turn produces a higher number of neighbors outside the Theiler exclusion window and therefore more points are effectively available to compute the cost function.
However, some problems may arise due to undersampling:
i) inside the embedding window we may not find enough observations to unfold the attractor,
ii) in the case of noncoherent dynamics, the sampling could be insufficient to describe the faster oscillations as discussed in \cite{Letellier:2009}, and
iii) the optimal value of $t_w$ can not be studied with the desired resolution.
From the description and analysis of the time series given in \cite{Hubner:1994} we can in this case eliminate possibilities i) and ii) above.

\begin{figure}[h]
\centering
\includegraphics[width=0.48\textwidth]{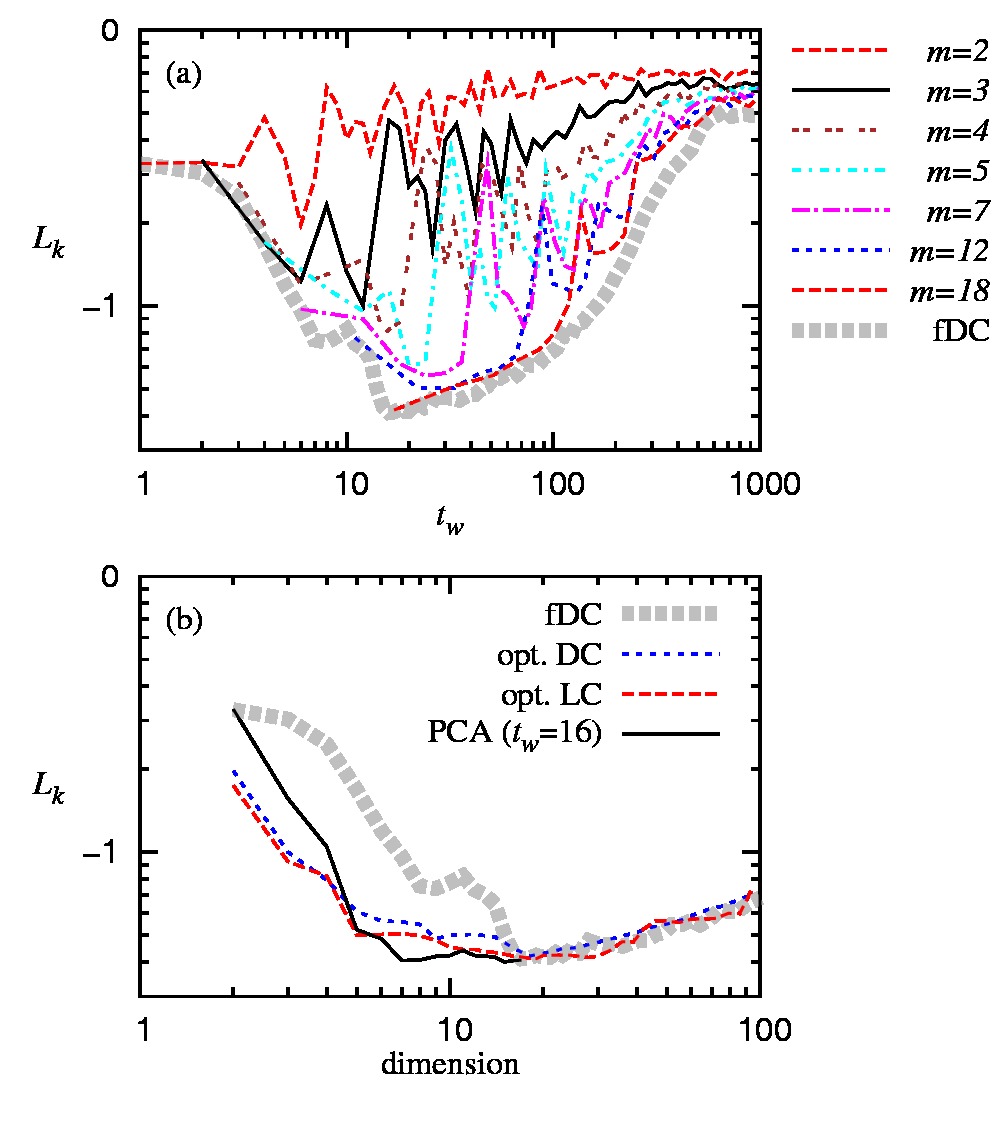}
\caption{(Color online). Evaluation of the cost function for different reconstruction strategies applied to the SFI laser time series. (a) $L_k$ vs.\ window size $t_w$ for $m$-dimensional DC reconstructions and for the fDC reconstruction (thick dashed gray line). (b) As a function of the dimension $m$, we plot i) $L_k$ for fDC (thick dashed gray line), ii) min$_{t_w}L_k$ for DC, iii) min$_{t_w}L_k$ for Legendre coordinate reconstructions, and iv) $L_k$ for PCA reconstructions. In the latter case, principal components are incorporated in decreasing order of their corresponding eigenvalues (full line).}
\label{fig:laser}
\end{figure}

Figure \ref{fig:laser}(a) shows the values of $L_k$ for DC reconstructions as a function of $t_w$ for different dimensions $m$.
The absolute minimum of $L_k$ occurs for $t_w=16$ and $m=17$, i.e.\ considering all delayed values inside the selected window (fDC).

As discussed in Sec.\ \ref{sec:noisy}, high dimensional DC reconstructions allow noise filtering when computing distances between vectors and this is probably the reason why the absolute minimum is found at $m=17$.
From \ref{fig:laser}(a) it can be argued that a reconstruction of dimension $m=5$ can be achieved with only a slight increase in the cost function.
To investigate whether this increase is relevant for reconstruction quality we considered the distribution of the local cost function in the same way as it was done in Fig.\ \ref{fig:color}.
We found (figure not shown) that this slight increase in $L_k$ is due to large increases of the local cost function at localized regions of the attractor which involve a small fraction of points (as it can also be observed in Fig.\ \ref{fig:color}(b)). This gives a meaning to the difference between $m=5$ and $m=17$ in the global cost function which can be then considered relevant.
As far as the time window width is concerned, we found that it approximately corresponds to the size of two characteristic periods.
We therefore see that, in contrast to previously considered systems, our analysis suggests that for this time series the irrelevance time is larger than the characteristic period.

In Fig.\ \ref{fig:laser}(b) we plot for each dimension $m$ the minimum value of $L_k$ over $t_w$ for the DC reconstructions of panel (a), and also for Legendre coordinates.
In the latter case the minimum of $L_k$ is reached at $t_w=19$ and $n=20$, which means that no dimension reduction is achieved.

To illustrate the versatility of the proposed approach, we also considered PCA coordinates in order to compare with the performance of the Legendre approach. More precisely, we applied PCA to the $t_w=16$ fDC reconstruction, which was the optimal DC reconstruction, and
computed $L_k$ for reconstructions of increasing dimension by sequentially incorporating the PCA components in decreasing order of their corresponding eigenvalue (full line profile in Fig.\ \ref{fig:laser}(b)).
After the 7th PCA component $L_k$ falls below the Legendre profile and reaches a minimum, thereby substantially reducing the dimensionality of the representation.

\begin{figure}
\centering
\includegraphics[width=0.48\textwidth]{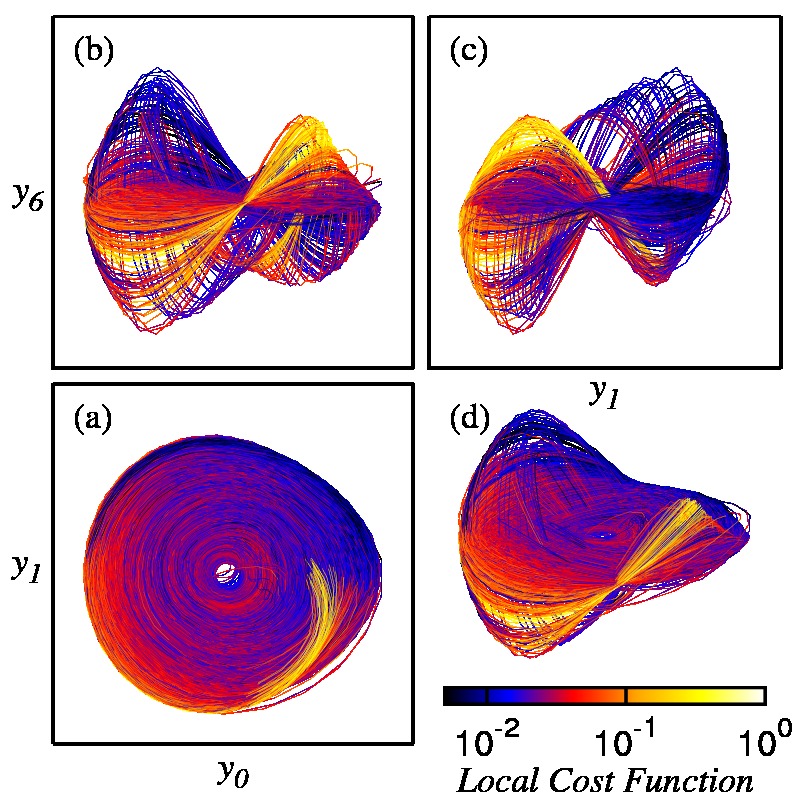}
\caption{(Color online). Views of a 3-dimensional reconstruction of the SFI laser attractor using PCA coordinates . (a) $y_1$ vs.\ $y_0$, (b) $y_6$ vs.\ $y_0$ and (c) $y_6$ vs.\ $y_1$. (d) View in perspective of the reconstructed attractor. The local cost function for the 7-dimensional PCA reconstruction is color-coded over the attractor and penalizes regions of high divergence. A spline interpolation was used to connect points in order to guide the eye for this undersampled time series.}
\label{fig:color_laser}
\end{figure}

Figure \ref{fig:color_laser} shows a 3-dimensional reconstruction using coordinates $y_0$, $y_1$ and $y_6$ obtained from PCA.
The reconstructed attractor exhibits a structure reminiscent of the R\"ossler attractor: a growing spiral in the $(y_0,y_1)$ plane with a reinjection of orbits at different radii.
Indeed, the equations describing the system dynamics are equivalent to Lorenz equations but the symmetric two-spiral structure of the Lorenz attractor is collapsed into one single spiral by the measurement process \cite{Hubner:1994}.
The values of the local cost function have been computed for the n=7 dimensional reconstruction and color-coded in the 3-dimensional projection of Fig.\ \ref{fig:color_laser}.
Notice the costly region corresponding to the reinjection of orbits into the spiral.
In contrast, in flatter regions of the spiral orbits are more predictable and show lower local objective function values.

\subsection{Non-uniform delay coordinate reconstructions}

We now report an exploration of the potential of this approach for the construction of non-uniform delay coordinate (nuDC) vectors, i.e.\ the case where consecutive delayed coordinates are not equidistant.
We briefly report a case study taken from Pecora {\it et al.}, who in \cite{Pecora:2007} introduced an iDC reconstruction method and used it to analyze a quasiperiodic, multiple time-scale time series of the $x$-coordinate of a orbit in a two-dimensional torus living in a 3-dimensional space.
The two frequencies are $\omega_1$ and $\omega_2$ with $\omega_2=2.5\pi\omega_1$ and the time series is sampled 32 times per fast cycle.

On one hand, applying a greedy search algorithm, Pecora {\it et al}.\ arrived at a 4-dimensional reconstruction with delay times $\tau_1=8$, $\tau_2=67$, and $\tau_3=75$.
The first delay $\tau_1=8$ captures the fast frequency, being exactly 1/4 of the corresponding period (this fraction is the time lag where autocorrelation vanishes for harmonic signals). However, $\tau_2=67$ slightly fails to capture the slow frequency (it should be $\tau_2=63$).

On the other hand, we exhaustively searched over the complete space of parameters $\{m,\tau_1,\tau_2,...,\tau_{(m-1)}\}$ for the minimum value of $L_k$.
According to our proposed measure $L_k$, the optimal iDC reconstruction is attained for parameter values $m=4$, $\tau_1=8$, $\tau_2=63$, and $\tau_3=71$.  As we can see, in this solution $\tau_1$ and $\tau_2$ exactly capture the fast and slow frequencies of this torus time series.

Finally, we used independent sets of random samples and ran bootstrapping experiments to compare the value of $L_k$ on the above reconstructions.  First, we found that the iDC reconstruction obtained by the approach here proposed is significantly better, in a statistical sense, than the one found by Pecora {\it et al}.  This result was expected by construction (except possibly the statistical significance).   
Secondly, we also found that according to our measure the iDC reconstruction obtained by Pecora {\it et al}.\ is in turn statistically significantly better than the best possible uniform DC reconstruction.

\section{Conclusions}\label{sec:conclusions}

In this work we considered the reconstruction problem as an optimization case, and proposed an objective function to guide the search for an optimal state space reconstruction.  This cost function is based on 1) the hypothesis of an underlying deterministic dynamics, 2) theoretical arguments on noise amplification, and 3) the idea of minimizing the complexity of the reconstruction.  It incorporates a novel irrelevance measure based on the characteristic distance to nearest neighbors in the reconstructed space.  The latter statistics captures in a simple and intuitive way attractor stretching ---a typical feature of overfolded reconstructions.

The proposed objective function can be evaluated on any reconstructed attractor, thereby enabling a direct comparison among different approaches: (uniform or non-uniform) delay vectors, PCA, Legendre coordinates, etc.  It can also be used to select the most appropriate parameters of a particular reconstruction strategy by searching for the absolute minimum of the advocated cost function.  For example, in the case of delay coordinates the search for the optimal delay time and embedding dimension can be automated by simply exploring the corresponding parameter space.  The absolute character of this search is in contrast with subjective choices of other methods in the literature, such as the value of a threshold to define false neighbors, or a tolerance limit to discriminate a noise-induced fluctuation from a true relative minimum.

Our approach has only two free parameters: the number of nearest neighbors $k$ and the prediction horizon $T_M$.  We have given a simple guidance to choose appropriate ranges for these parameters, where results depend mildly on the particular configuration and the method returns a robust output.  Code implementing the proposed method is freely available (see Appendix \ref{ap:algorithm}).

We applied the proposed method for the analysis of several synthetic and experimental times series.  Among the latter we considered field measurements from an experimental realization of Chua's circuit
and a far-infrared laser taken from the Santa Fe Institute time series prediction competition.  In particular, we used the latter examples to demonstrate the ability of the proposed approach to handle different types of reconstructions, which we believe to be a distinctive and powerful feature of this method.  In all cases we found a well defined absolute minimum of the objective function.

In the particular case of delay coordinate embeddings we found the interesting result that the time span of the optimal reconstruction window is not necessarily related to the characteristic period of the time series under consideration.
The results obtained for the
SFI laser time series suggest that measurements from more than one cycle in the past are relevant for the reconstruction of the system state.

As a final remark, we would like to notice that in the present work we have not proposed a new reconstruction strategy but a new methodology to measure the quality of a reconstruction. From the case studies analyzed in this work we conclude that none of the considered reconstruction techniques (delay coordinates, PCA, Legendre coordinates) is universally optimal.  From a practitioner's point of view, the proposed cost function is therefore useful to assess which is the most appropriate reconstruction method for the particular time series under study.

\acknowledgments

We thank two anonymous referees for valuable input that improved an earlier version of this manuscript.  G.L.G. and L.C.U. were partially supported by ANPCyT grant PICT-2008-0237.

\appendix

\section{Forecasting accuracy comparison}\label{ap:prediction}

As an objective validation of the proposed approach we compared the forecasting accuracy of local linear models in reconstruction spaces obtained with the standard approach (Mutual Information + False Nearest Neighbors) and our method.
For the Mackey-Glass time series the standard embedding method gives $m=4$ and $\tau=22$ (an embedding window of size $t_w=66$) while the proposed method gives $m=4$ and $\tau=10$ (i.e., a smaller window size $t_w=30$). For the $x$-coordinate of the R\"ossler time series the standard embedding space is $m=3$, $\tau=11$ ($t_w=22$) while our approach gives $m=3$, $\tau=5$ ($t_w=10$). Finally, for the $x$-coordinate of the Lorenz system the standard method yields $m=3$, $\tau=16$ ($t_w=32$) and the proposed one $m=3$, $\tau=6$ ($t_w=12$).

In these reconstruction spaces, for a given vector $y(t)$ in an independent {\it test set} (not used to determine the reconstruction parameters) the prediction algorithm identifies the first $k$ neighbors among the $N$ available in the {\it training set} (also used to determine the reconstruction parameters) and locally fits a linear model.

A comparison of the normalized root mean squared prediction error on data sets from the Mackey-Glass, R\"ossler, and Lorenz systems (Appendices \ref{ap:mcgl}, \ref{ap:rossler} and \ref{ap:lorenz} respectively) is depicted in Fig.\ \ref{fig:validation}.
In the upper panels we show the prediction error as a function of the number of neighbors $k$ (more precisely, as a function of the fraction $k/N$ where $N$ is the size of the training set).  We fixed the horizon $T$ at the first maximum of the space-time separation plot as discussed in Section \ref{sec:Gao}.
In the lower panels of Fig.\ \ref{fig:validation} we show the prediction error as a function of the horizon $T$ (expressed in units of the characteristic period of each time series) for a fixed, non-optimized, arbitrarily chosen number of neighbors $k=15$.

\section{Mackey-Glass dataset}\label{ap:mcgl}

We considered the Mackey-Glass equation \cite{Mackey:1997}:
\begin{equation*}
\frac{dx}{dt} = \frac{ax(t-\overline{\tau})}{1+x^c(t-\overline{\tau})}-bx \nonumber
\end{equation*}
with parameters $a=0.2$, $b=0.1$, $c=10$, and $\overline{\tau} = 17$. Using as initial conditions $x(t<0)=0$ and $x(t=0)=1.2$, after integrating this infinite-dimensional differential equation and sampling every $\delta t = 0.5$ we kept a 10,000-element time series following
the first 2,000 iterations which were discarded as a transient.  We also computed a 50,000-element continuation used only as a test set in the prediction task, and the same was done for R\"ossler and Lorenz below.  

\section{R\"ossler dataset}\label{ap:rossler}

For the R\"ossler system \cite{Roessler:1976} we integrated the equations
\begin{eqnarray*}
dx/dt &=& -(y+z), \\
dy/dt &=& x + \alpha y, \\
dz/dt &=& \beta + z(x-\gamma)
\end{eqnarray*}
with parameters $\alpha = 0.15$, $\beta = 0.2$, and $\gamma = 10$. As initial conditions we used $x=y=z=1$, we then employed a 4th order Runge-Kutta numerical integration method, sampled the $x$-coordinate of the obtained trajectory with a step $\delta t = 0.125$, and finally discarded the first 8,000 data points keeping a time series of length 10,000, plus an extra 50,000 for testing.

\section{Lorenz dataset}\label{ap:lorenz}

The Lorenz system \cite{Lorenz:1963} is described by the equations
\begin{eqnarray*}
dx/dt &=& \sigma (y-x), \\
dy/dt &=& x(\rho -z) - y, \\
dz/dt &=& xy - \beta z
\end{eqnarray*}
with parameters $\sigma=10$, $\rho=28$, and $\beta=8/3$. We initialized the system at $x=y=1$, $z=50$ and integrated with a 4th order Runge-Kutta procedure with step $\delta t = 0.01$. After a transient of 1,000 data points we kept 10,000 samples from the $x$-coordinate of this flow, and additional 50,000 for independent testing.

\section{Implementation}\label{ap:algorithm}

The computation of the cost function requires to perform nearest neighbor searches in high dimensional spaces.
To this end we use a box-assisted algorithm for efficient neighbor searching \cite{Kantz:1997}.
This algorithm is an extension of the False Nearest Neighbor (FNN) algorithm of the TISEAN package \cite{Hegger:1999}.
This extension allows the search of $k$ nearest neighbors of $\bar{x}$ to define the set $\mathcal{U}_k(\bar{x})$ where none of the neighbors belong to the same Theiler window.

An  implementation in C code of the method proposed in this work can be downloaded from \cite{www:algorithm}.
The program computes the global and local cost function for DC, fDC and Legendre coordinates, which are internally implemented.
Alternatively, it also allows loading reconstructions from external files, and then computes corresponding local or global cost function values. We also provide scripts to reproduce Figs.\ \ref{fig:example_mcgl}, \ref{fig:color}, \ref{fig:dsil}(b) and \ref{fig:color_dsil}.


\begin{thebibliography}{}
\bibitem{comment:resolution} In the case of low resolution experimental time series, Eq.\ \ref{eq:sigma_k} can often produce a division by zero given that close states are mapped onto identical measurement values. To avoid this, we added uniform noise below the resolution level to the time series in order to satisfy the hypothesis of $x$ being generated by a smooth measurement function plus noise. This procedure does not reduce the signal to noise ratio.
\end{thebibliography}
\end{document}